\documentclass[amsmath,amssymbs,aps,twocolumn]{revtex4-2}

\usepackage{graphicx}
\usepackage{amsmath,amsfonts,amssymb}
\usepackage{epsfig}
\usepackage{wrapfig}
\usepackage{bbm}
\usepackage[usenames]{color}
\usepackage{array}
\usepackage{times}
\usepackage{float}
\usepackage{epstopdf}

\begin{document}
	
	%\preprint{APS/123-QED}
	
	\title{Reconfigurable Non-Hermitian Soliton Combs using Dissipative Couplings and Topological Windings}
	
	\author{Seyed Danial Hashemi}
	
	\author{Sunil Mittal}
	\email[Email: ]{s.mittal@northeastern.edu}
	
	\affiliation{Department of Electrical and Computer Engineering, Northeastern University, Boston, MA 02115, USA}
	\affiliation{Institute for NanoSystems Innovation, Northeastern University, Boston, MA, 02115, USA}
	%\date{\today}

\begin{abstract} 
The emergence of dissipative Kerr solitons (DKS) in nonlinear resonators has revolutionized the generation of on-chip coherent optical frequency combs. The formation of DKS in conventional single resonators hinges on balancing the resonator dissipation against the parametric gain and balancing the resonator dispersion against the resonance frequency shifts introduced by the Kerr nonlinearity. Here, we theoretically introduce a new class of non-Hermitian soliton combs that are enabled by engineering the dissipation and dispersion of a coupled resonator array with nonreciprocal couplings. We show that these non-Hermitian soliton combs allow unprecedented post-fabrication agile reconfigurability of the soliton comb spectrum, where the number of comb lines, as well as their frequency spacing, can be drastically tuned by simply tuning the hopping phases between resonators. Such reconfigurable non-Hermitian combs generated using coupled resonator arrays could enable new functionalities for a multitude of comb applications. 
\end{abstract}

\maketitle
%%%%%%%%%%%%%%%%%%%%%%%%%%%%%%%%%%%%%%%%%%%%%%%%%%%%%%%%%%%%%%%%%%%%%%%%%%%%%
% Introduction
%%%%%%%%%%%%%%%%%%%%%%%%%%%%%%%%%%%%%%%%%%%%%%%%%%%%%%%%%%%%%%%%%%%%%%%%%%%%
\section*{Introduction}

Dissipative Kerr solitons (DKS) are pulses of light that preserve their shape when propagating in a dispersive and dissipative medium with Kerr nonlinearity. DKS emerge naturally while generating coherent optical frequency combs in nonlinear resonators when the dissipation of the resonator is compensated for by the nonlinear gain and the dispersion of the resonator is compensated for by the dispersion induced by nonlinearity \cite{Cundiff2003, Kippenberg2011, Kippenberg2018, Pasquazi2018, Gaeta2019}. Although engineering the resonator waveguide geometry is the most commonly used approach \cite{Kippenberg2018, Anderson2022}, recently techniques such as embedding photonic crystals or inverse-designed reflectors in ring resonators \cite{Yu2021, Yu2022, Lucas2023, Yang2023}, and coupled resonator systems in the form of photonic molecules \cite{Miller2015, Kim2017, Jang2018, Vasco2019, Helgason2021, Tikan2021, Yuan2023, Helgason2023}, have been explored to engineer the resonator dispersion and, thereby, the spatio-temporal behavior of Kerr soliton combs. More recently, large arrays of coupled resonators hosting topological edge states with linear dispersion have been shown to generate novel nested solitons and nested frequency combs that are not achievable using single resonators \cite{Mittal2021b, Flower2024, Tusnin2023, Hashemi2024}. Nevertheless, analogous advances in engineering the dissipation of the resonators to control the formation of soliton combs have remained largely elusive. 

In parallel, recent advances in the field of non-Hermitian physics of open quantum systems have led to the development of a new paradigm that uses dissipation to engineer system behavior \cite{Gong2018, Ashida2020, Okuma2020, Kawabata2019, Okuma2023, El-Ganainy2018, Miri2019,   Nasari2023, Li2023, Wang2023, Fang2017, Metelmann2015}. In particular, non-Hermitian model systems, such as the Hatano-Nelson model with nonreciprocal couplings and open boundary conditions, exhibit the skin effect where all modes are exponentially localized at the boundary of the lattice \cite{Gong2018, Ashida2020, Okuma2020, Weidemann2020, Leefmans2022, Ding2022, Wang2023}. In systems with periodic boundaries, this non-Hermitian topology manifests itself as topologically non-trivial windings in the two-dimensional (2D) complex energy plane spanned by the real and imaginary eigenvalues \cite{Wang2021, Wang2021b, Ding2022, Wang2023}. These windings reveal the intertwined nature of the dispersion (real eigenvalues) and dissipation (imaginary eigenvalues) of the non-Hermitian system where some modes experience higher dissipation than others. 

Here, we theoretically introduce a new paradigm of non-Hermitian optical frequency combs where we use engineered dissipation in a nonlinear system to enable the formation of Kerr solitons. Specifically, we consider an array of nonlinear ring resonators that are coupled using non-Hermitian dissipative couplings and non-zero hopping phases (a synthetic magnetic flux). By engineering the couplings strengths and hopping phases, we engineer the topological windings, and equivalently, the dispersion and dissipation of the supermodes of this array. We show that this simultaneous dissipation-dispersion engineering allows the formation of nested Kerr solitons and nested optical frequency combs \cite{Mittal2021b, Tusnin2023, Flower2024} in the ring resonator array. These nested combs feature a comb-in-a-comb structure with comb lines repeating at two very different frequency scales.  In the absence of such non-Hermitian engineering, this array will not support the formation of stable solitons because of undesired nonlinearity-induced mixing between the supermodes. Even more so, we show that non-Hermitian engineering allows unprecedented post-fabrication reconfigurability of the generated nested combs, where both the comb line spacing and the number of comb lines can be dynamically reconfigured by tuning the hopping phases (for example, using thermal heaters). Such extreme post-fabrication reconfigurability is not achievable using conventional single-resonator Kerr combs, where the comb line spacing is set by the free spectral range of the resonator.

Our results could lead to transformative changes in the applications of combs that span radio frequency signal synthesis and processing, frequency and time metrology, spectroscopy, and light detection and ranging, and enable, for example, a reconfigurable platform for optical-comb-based RF signal synthesis and processing \cite{Chembo2016, Pasquazi2018}. On a more fundamental level, our results add the rich physics of frequency combs and temporal soliton formation to growing field of nonlinear non-Hermitian systems, that have recently led to the demonstration of, for example, non-Hermitian spatial solitons \cite{Pernet2022}, nonlinearity-controlled topological edge states and phase transitions\cite{Xia2021,  Dai2023, Reisenbauer2024, Li2023}, and non-Hermitian lasers \cite{Liu2022, Leefmans2024}.

%%%%%%%%%%%%%%%%%%%%%%%%%%%%%%%%%%%%%%%%%%%%%%%%%%%%%%%%%%%%%%%%%%%%%%%%%%%%%
% Figure 1
%%%%%%%%%%%%%%%%%%%%%%%%%%%%%%%%%%%%%%%%%%%%%%%%%%%%%%%%%%%%%%%%%%%%%%%%%%%%%
\begin{figure*}[ht]
 \centering
 \includegraphics[width=0.98\textwidth]{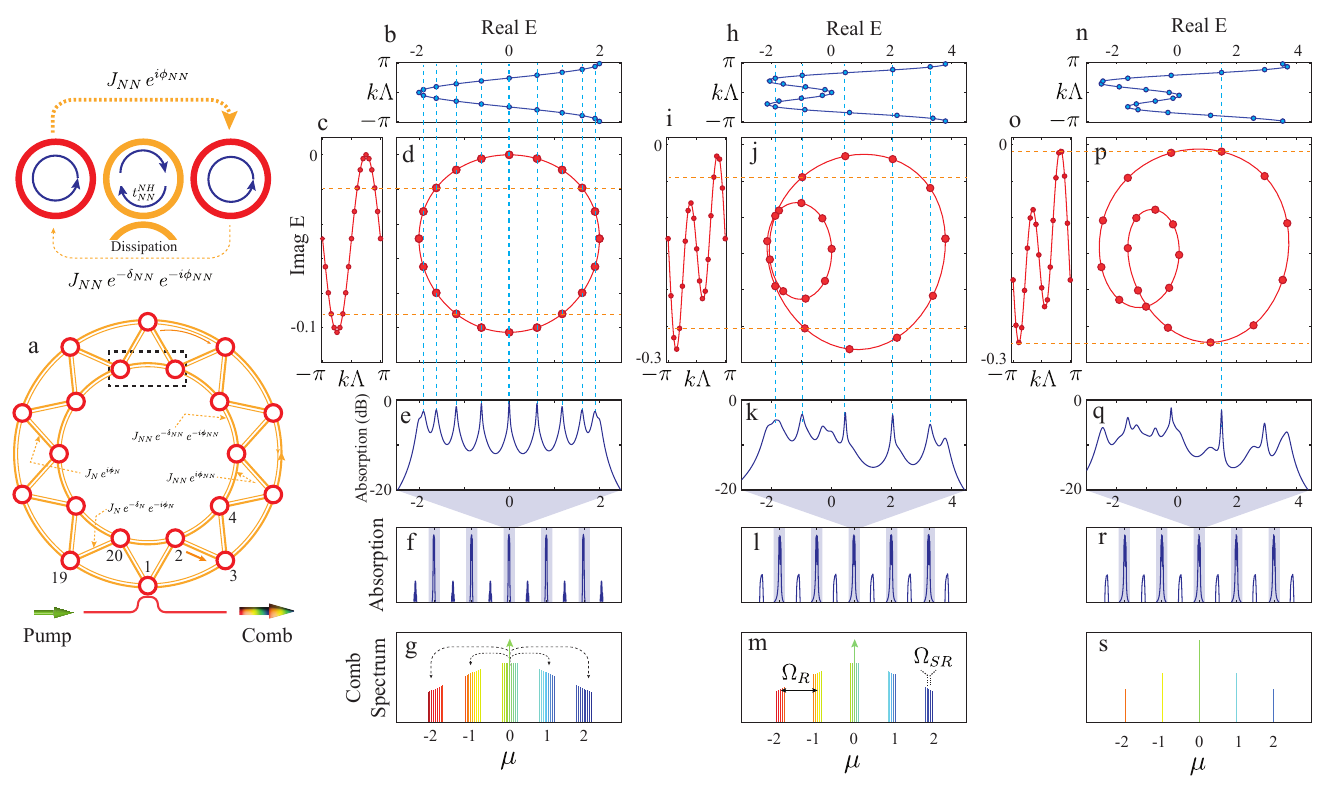}
 \caption{\textbf{Reconfigurable non-Hermitian soliton combs.}  \textbf{a.} Schematic of the non-Hermitian array with site rings (shaded red and numbered as shown), dissipatively coupled to the nearest and next-nearest neighbors via link rings (shaded orange). The zoom-in shows an external waveguide coupled to the link ring that creates the dissipative couplings. The array is coupled to an input-output waveguide to inject pump light and out-couple generated comb. \textbf{b,c.} Real and imaginary part of energy (frequency) eigenvalues, for a system with only nearest-neighbor couplings, as a function of Bloch momentum $k\Lambda$ in the array. Here, $J_N = 1, ~J_{NN} = 0, ~t^{NH}_{N} = 0.95, ~\phi_{N} ~= ~\phi_{NN} ~= ~0$. \textbf{d.} real and imaginary parts of eigenvalues plotted on a 2D complex plane show topological windings, here in the form of a circle. \textbf{e.} Simulated absorption (1-transmission) spectrum of the linear array near a single FSR of the site rings. \textbf{f.} Absorption spectrum over multiple FSRs. The spectrum in \textbf{e.} is a zoom-in near one FSR of the site rings. \textbf{g.} Schematic of the generated nested comb spectrum with nine supermodes oscillating at each FSR (labeled by $\mu$). \textbf{h-m.} Corresponding results for a lattice with both nearest and next-nearest neighbor couplings, with $J_N = J_{NN} = 1, ~t^{NH}_{N} = 0.95, ~t^{NH}_{NN} = 0.9, ~\phi_{N} = 0,  ~\phi_{NN} = 0.50625 (2\pi)$. We now observe a double winding in the 2D complex energy plane and an engineered absorption spectrum, resulting in a nested comb with only five supermodes oscillating in each FSR. \textbf{n-s} Corresponding results for a lattice with a different hopping phase, such that $J_N = J_{NN} = 1, ~t^{NH}_{N} = 0.95, ~t^{NH}_{NN} = 0.9, ~\phi_{N} = 0,  ~\phi_{NN} = 0.41667 (2\pi)$. In this case, the generated comb spectrum shows the oscillation of only a single supermode in each FSR $\mu$.}
 \label{fig:1}
\end{figure*}
%%%%%%%%%%%%%%%%%%%%%%%%%%%%%%%%%%%%%%%%%%%%%%%%%%%%%%%%%%%%%%%%%%%%%%%%%%%%%

Our system consists of a one-dimensional (1D) array of site ring resonators arranged as a super-ring with periodic boundary conditions (Fig.1a). The site rings are coupled to their nearest and next-nearest neighbors using another set of rings called the link rings \cite{Hafezi2011, Hafezi2013}. The resonance frequencies of the link rings are shifted from those of the site rings by one-half free-spectral range (FSR), for example, by decreasing their length. The use of an additional waveguide coupled to the link rings allows us to introduce non-Hermitian dissipative couplings such that the photons experience a direction-dependent loss when hopping between neighboring site rings (Fig.1a). In this configuration, the effective coupling rate for photons hopping towards the right neighboring site rings $\left(J_{N}, ~ J_{NN} \right)$ differs from those hopping towards the left $\left(J_{N}~e^{-\delta_{N}}, ~ J_{NN}~e^{-\delta_{NN}} \right)$. The nonreciprocal part of the couplings $e^{-\delta_{N}}, e^{-\delta_{NN}}$ is dictated by the transmission coefficients $t^{NH}_{N}$ and $t^{NH}_{NN}$ of the dissipative coupling regions of the link rings (Fig.1a). Therefore, this system is non-Hermitian when $\delta_{N} > 0$ and/or $\delta_{NN} > 0$, or equivalently, when $t^{NH}_N < 1$ and/or $t^{NH}_{NN} < 1$. We note that the effective coupling strength in either direction can be decreased by appropriately choosing the position (up or down) of the coupling waveguide that introduces direction-dependent loss. Appropriately shifting the link rings also allows us to introduce a direction-dependent phase $\left( \pm \phi_{N}, ~\pm \phi_{NN}\right)$ when the photons hop between neighboring site rings \cite{Hafezi2011, Hafezi2013}. The simultaneous presence of direction-dependent hopping phases and hopping loss effectively creates both a real and an imaginary synthetic magnetic field for photons. The array is coupled to an input-output waveguide, at one of the site rings, which allows injection of a pump laser and out-coupling of the generated frequency comb. 

We use the transfer-matrix formalism to calculate the frequency (energy) eigenvalues of this periodic, linear system as a function of the Bloch momentum $k$. The eigenvalues are calculated near a given longitudinal mode of single-ring resonators such that $\omega_{0,\mu}$ is the resonance frequency of the longitudinal mode labeled $\mu$, and we set $\omega_{0,0} = 0$. The eigenvalues are presented in Fig.1 for different choices of hopping strengths and hopping phases. In general, we note that the eigenvalues are complex. The real part of these eigenvalues represents the resonance frequencies, and therefore the dispersion of the supermodes. Their imaginary part, which is always $\leq 0$, represents the effective dissipation of the supermodes. More importantly, we find that some of the supermodes experience much higher dissipation compared to others. 

To reveal the interplay between the dispersion and dissipation of this non-Hermitian periodic system, we plot the real and imaginary eigenvalues on a 2D complex plane, as shown in Fig.1 d, j, p. In particular, we observe the emergence of topologically non-trivial closed-loop windings that govern the symmetries of the system with respect to the dispersion and dissipation of its supermodes \cite{Wang2021, Wang2021b, Ding2022, Wang2023}. For example, for a system with only nearest-neighbor couplings $\left(J_{NN} = 0\right)$ and no hopping phases $\left(\phi_{N} = \phi_{NN} = 0 \right)$, the topological winding is in the form of a circle such that the resonance frequencies (real eigenvalues) of the supermodes are located symmetrically around the ring resonance frequency $\omega_{0}$ (which we set to zero). Furthermore, the supermodes are two-fold degenerate in their real part. The presence of dissipative couplings breaks this degeneracy in their imaginary part, such that the supermode with momentum $-k$ experiences higher dissipation compared to the one with momentum $+k$ (Fig.1c). By introducing next-nearest neighbor couplings and hopping phases, we can introduce higher-order topological windings that can break these symmetries.

This eigenvalue structure is also manifested in the absorption (1-transmission) spectrum of the array, where we consider injecting light at the input and measuring the transmission at the output of the input-output waveguide. In particular, the frequencies at which we observe peaks in the absorption spectrum correspond to the real part of the eigenvalues. Similarly, we observe higher absorption in the array, through the input-output waveguide, for supermodes that have lower loss (or lower absolute imaginary eigenvalues). Evidently, by tuning hopping strengths, hopping phases, and the dissipation in the link rings, we can engineer the windings or, equivalently, both the dispersion and the dissipation of the supermodes and their absorption spectra.

To generate frequency combs in the lattice, we consider a continuous-wave pump coupled to the lattice via the input-output waveguide. The pump generates the frequency comb via the nonlinear four-wave mixing (FWM) process. The number of supermodes participating in the generated comb spectra is controlled by both their real and imaginary eigenvalues. In particular, the real part of the eigenvalues dictates the supermode dispersion or energy conservation in the FWM process. The imaginary part of the eigenvalues dictates the balance between the loss introduced by dissipative couplings and the gain introduced by the FWM process. Therefore, only the supermodes that exhibit low loss (higher absorption) and are located approximately symmetrically around the pumped supermode contribute to the formation of coherent soliton combs. In the following, we will show that non-Hermitian engineering using dissipative couplings enables the formation of reconfigurable Kerr solitons and coherent optical frequency combs by selectively dissipating a set of supermodes and, thereby, suppressing undesired nonlinear mixing between the supermodes. 

To simulate the generation of frequency combs, we use the Ikeda map formalism as detailed in the Supplementary Information \cite{Ikeda1979, Hansson2015, Hansson2016}. We emphasize that the commonly used Lugiato-Lefever formalism, which relies on the single-mode approximation and the effective Hamiltonian formalism, is not well suited for these simulations \cite{Little1997, Chembo2013, Hansson2016}. This is because of the spurious gain terms introduced by the non-Hermitian couplings when using the single-mode approximation (see Supplementary Materials). For our Ikeda map simulations, we use dimensionless normalized parameters such that the group velocity $v_{g}$, the length of the site rings $L_{R}$, and their free spectral range $\Omega_{R}/\left(2\pi \right)$ (FSR, in frequency units) are normalized to one. The coupling strength $J = 0.01 ~\Omega_{R}$, the loss rate of individual rings $\kappa_{in} = 0.01 J$. We assume an anomalous dispersion for the ring resonators, represented by the parameter $D_{2} = 5 \times 10^{-6} ~\Omega_{R}$. We do not make any assumptions regarding the dispersion or dissipation of the supermodes.\\

%%%%%%%%%%%%%%%%%%%%%%%%%%%%%%%%%%%%%%%%%%%%%%%%%%%%%%%%%%%%%%%%%%%%%%%%%%%%%
% Results
%%%%%%%%%%%%%%%%%%%%%%%%%%%%%%%%%%%%%%%%%%%%%%%%%%%%%%%%%%%%%%%%%%%%%%%%%%%%

\section*{Results}

\noindent
\textbf{Comb in lattices with only nearest-neighbor couplings}\\
To demonstrate the use of non-Hermitian dissipative couplings to control the formation of soliton combs, in Fig.2, we first present results for the super-ring with $N = 20$ site rings, with only the nearest neighbor couplings $J_{N} = J$, and all hopping phases set to zero ($\phi_{N} = \phi_{NN} = 0$). As discussed earlier, without dissipative couplings, such a system exhibits a two-fold degeneracy in its eigenvalues (Fig.1b, Fig.2a). One supermode of this pair circulates the super-ring in the clockwise direction and the other in the counter-clockwise direction. This degeneracy is also evident in the absorption spectrum of the array, where only $\sim N/2$ modes are visible as absorption peaks (Fig.2b). For a Hermitian system, all the eigenvalues are real, and it has a trivial topology (a line). Pumping the Hermitian super-ring at a single lattice site excites an equal superposition of both modes, and this competition hinders the formation of coherent soliton combs in the Hermitian super-ring array. 

%%%%%%%%%%%%%%%%%%%%%%%%%%%%%%%%%%%%%%%%%%%%%%%%%%%%%%%%%%%%%%%%%%%%%%%%%%%%%
% Figure 2
%%%%%%%%%%%%%%%%%%%%%%%%%%%%%%%%%%%%%%%%%%%%%%%%%%%%%%%%%%%%%%%%%%%%%%%%%%%%%
\begin{figure*}
 \centering
 \includegraphics[width=0.98\textwidth]{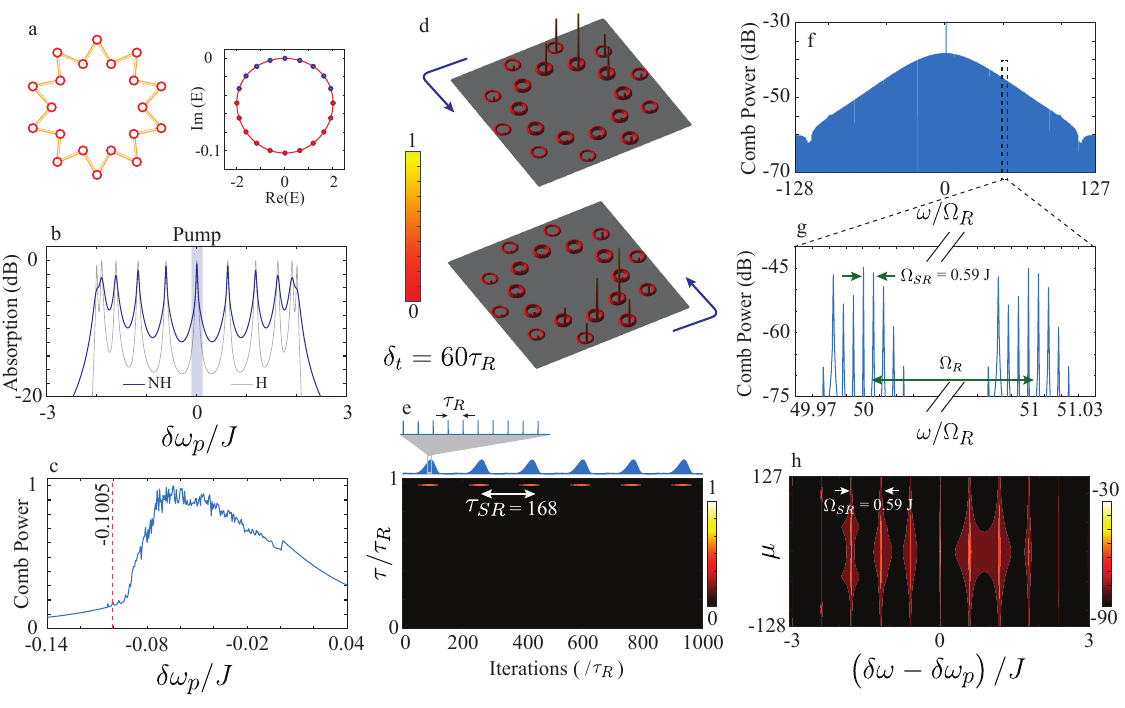}
\caption{\textbf{Soliton combs in resonator arrays with only nearest neighbor couplings.} \textbf{a.} Schematic of the array and the resultant winding in the form of a circle. The blue shaded dots on the winding curve indicate supermodes that contribute to the nested comb formation. \textbf{b.} Linear absorption (1-transmission) spectrum of the array, for Hermitian and non-Hermitian lattices. The pumped mode is shaded blue. \textbf{c.} Comb power as a function of pump frequency. The pump frequency where we observe nested solitons is indicated by the dashed line. \textbf{d.} Soliton intensity distribution and its circulation around the array in the CCW direction. For clarity, the plot only shows the site rings. \textbf{e.} Temporal output showing bursts of pulses, with each burst separated by the round-trip time $\tau_{SR} \simeq 168 \tau_{R}$ of the super-ring. Within each burst, the light pulses are separated by the round-trip time $\tau_{R}$ of the single rings. \textbf{f.} Generated comb spectrum,  and \textbf{g.} zoom-in of the spectrum around FSRs $\mu = 50$ and $\mu = 51$. We observe the formation of a nested comb with the oscillation of 9 supermodes in each FSR (similar to the schematic of Fig. 1g). The FSR of the sub-comb $\Omega_{SR} = 1/\tau_{SR} = 0.59 J = 0.0059 \Omega_{R}$. \textbf{h.} Comb spectrum re-organized as a function of fast $\left(\mu \right)$ and slow $\delta\omega$ frequency. The nested comb lines are straight, indicating cancellation of linear and nonlinear dispersion, and formation of a soliton nested comb.} 
\label{fig:2}
\end{figure*}
%%%%%%%%%%%%%%%%%%%%%%%%%%%%%%%%%%%%%%%%%%%%%%%%%%%%%%%%%%%%%%%%%%%%%%%%%%%%%

Nevertheless, as evident from Fig.2a, the introduction of dissipative couplings, with $t^{NH}_{N} = 0.95$ introduces non-trivial topological winding in the form of a circle. Although the supermodes stay doubly degenerate in their real eigenvalues, dissipative couplings break their degeneracy in the imaginary eigenvalues and suppress mixing between the CW and CCW propagating supermodes. For our choice of circulation direction in the site rings and the dissipative waveguide coupling in the link rings (Fig.1a), supermodes with momentum $-k$, traveling in the CW direction around the super-ring, have higher loss than those traveling along the positive $k$ (CCW) direction. The absorption spectrum of the non-Hermitian lattice is shown in Fig.2b, where we observe an increase in the linewidth of the supermodes because of dissipative couplings. We have also included intra-resonator dissipation, with rate $\kappa_{in} = 0.01 J$, in both the site and the link rings.

To generate coherent soliton combs, we pump the supermode at the center of the absorption spectrum shown in Fig.2b, and analyze the spatio-temporal and spectral response of the array. When we tune the amplitude of the normalized pump field $E_{in} = 0.021$ and the pump frequency detuning $\delta\omega_{p} = \omega_{p} - \omega_{0,0} = -0.10005 ~J $, in the spatio-temporal domain (Fig. 2d), we observe the formation of coherent nested solitons in the lattice, similar to those observed in 2D topological ring resonator arrays \cite{Mittal2021, Flower2024, Tusnin2023}. In this state, the super-ring hosts a single super-soliton - a set of a few consecutive site rings, each of which also hosts a single soliton. More importantly, the solitons within the site rings self-organize such that their position within the single rings is always the same; that is, they are phase-locked. As time evolves, this nested soliton structure circulates the lattice in the CCW direction but without losing its coherence (also see Movie S1). We do not observe any propagation in the CW direction. This indicates that none of the CW propagating modes participate in the comb formation because the nonlinear FWM gain is not sufficient to overcome the higher loss introduced by dissipative couplings.

This formation of a nested soliton in the lattice is manifested at the output of the lattice as the generation of nested pulses or bursts of light pulses (inset of Fig. 2e). The bursts are separated in time by the round-trip time $\tau_{SR} \sim 168 \tau_{R}$ of the super-ring resonator. Within each burst, the pulses are separated by the round trip time $\tau_{R}$ of the single rings. The temporal output of the lattice, presented in Fig.2e as a function of the fast time $\tau = \left(0,\tau_{R} \right)$ and the slow time (number of round trips or iterations), shows that pulses always appear at the same fast time $\tau$. This highlights the phase-locked nature of the solitons within individual rings (Fig.2d).

The generated frequency comb spectrum for this array is shown in Fig.2h. The intensity profile across the comb lines is smooth, which indicates the formation of a coherent soliton comb. In Fig.2g, we plot the comb spectrum zoomed in on two FSRs $\left(\mu = 50, 51 \right)$. We clearly observe the formation of a nested comb (Fig.2f-h) where a set of nine supermodes oscillates at each FSR $\mu$. The different sets of oscillating supermodes are spaced by the FSR $\Omega_{R}$ of the single rings. The frequency spacing between consecutive supermodes within a single set is much smaller, such that $\Omega_{SR} = 1/\tau_{SR} \simeq 0.59 ~J = 0.0059 \Omega_{R}$. We note that the number of oscillating modes (here nine) is dictated by the dispersion and dissipation of the array as depicted by its topological winding, where we find a set of nine modes that have a lower loss (indicated by dashed lines in Fig.1d) and are located symmetrically around the pumped mode. 

To further highlight the soliton nature of this comb, in Fig.2h, we plot the comb spectrum as a function of the longitudinal mode index $\mu$ of the single rings (which we refer to as the fast frequency) and the pump-normalized slow frequency $\delta\omega - \delta\omega_{p} =  \left(\omega - \omega_{0,\mu}\right) - \left(\omega_{p} - \omega_{0,0} \right)$ which reveals the oscillating supermodes within each FSR. Here $\omega_{0,\mu}$ is the resonance frequency of $\mu^{\text{th}}$ FSR (of single rings), and $\omega_{0,0}$ with $\mu = 0$ is that for the pumped FSR. $\omega$ corresponds to the frequency of an oscillating supermode. We find that the nested comb lines do not show any curvature, which clearly indicates the cancellation of linear and nonlinear dispersion and the formation of coherent solitons. This spatio-temporal response and frequency comb spectrum of nested soliton combs can be contrasted with that of chaotic combs (see Supplementary Fig.S5), where we clearly observe an absence of smooth comb spectrum, the underlying linear dispersion of the cavity manifested as the curvature of the generated comb lines, and an absence of stable spatial or temporal pattern formation. \\

\noindent
\textbf{Combs in lattices with nearest and next-nearest neighbor couplings}\\
Next, we demonstrate the use of higher-order non-Hermitian topological windings to engineer the comb spectrum. Specifically, we introduce next-nearest-neighbor couplings $J_{NN}$ and set their strength such that $J_{NN} ~=  ~J_{N} ~= ~J$. We chose non-reciprocal couplings for both nearest and next-nearest neighbors such that $t^{NH}_{N} = 0.95$ and $t^{NH}_{NN} = 0.9$. We also introduce next-nearest-neighbor hopping phases $\phi_{NN} = 0.50625 (2\pi)$. As shown in Fig.3a, this system exhibits a non-Hermitian topology with double winding. The absorption spectrum for this system is shown in Fig.3b, and is compared to that of a Hermitian system. We note that because of next-nearest-neighbor couplings, the range of real eigenvalues $\sim \left(-2.5J, 4J \right)$ is much larger compared to the super-ring with only nearest-neighbor couplings where this range was $\sim \left(-2J, 2J \right)$.  More importantly, the presence of the hopping phase $\phi_{NN}$ lifts the degeneracy in the real eigenvalues, and the Hermitian spectrum exhibits $N = 20$ peaks. However, the eigenvalues are non-uniformly distributed across the frequency spectrum. This non-uniformity can also be observed in the asymmetric location of the smaller winding loop in Fig.3a. This skewed dispersion and undesired couplings between closely spaced counter-propagating supermodes make it challenging to achieve soliton combs in the Hermitian lattice. By introducing dissipative couplings, we engineer the absorption spectrum such that only five supermodes, nearly equally spaced in frequency, exhibit lower dissipation and, consequently, higher absorption (highlighted by dashed lines in Fig.1j and blue shaded dots in Fig.3a). 

%%%%%%%%%%%%%%%%%%%%%%%%%%%%%%%%%%%%%%%%%%%%%%%%%%%%%%%%%%%%%%%%%%%%%%%%%%%%%
% Figure 3
%%%%%%%%%%%%%%%%%%%%%%%%%%%%%%%%%%%%%%%%%%%%%%%%%%%%%%%%%%%%%%%%%%%%%%%%%%%%%
\begin{figure*}
 \centering
 \includegraphics[width=0.98\textwidth]{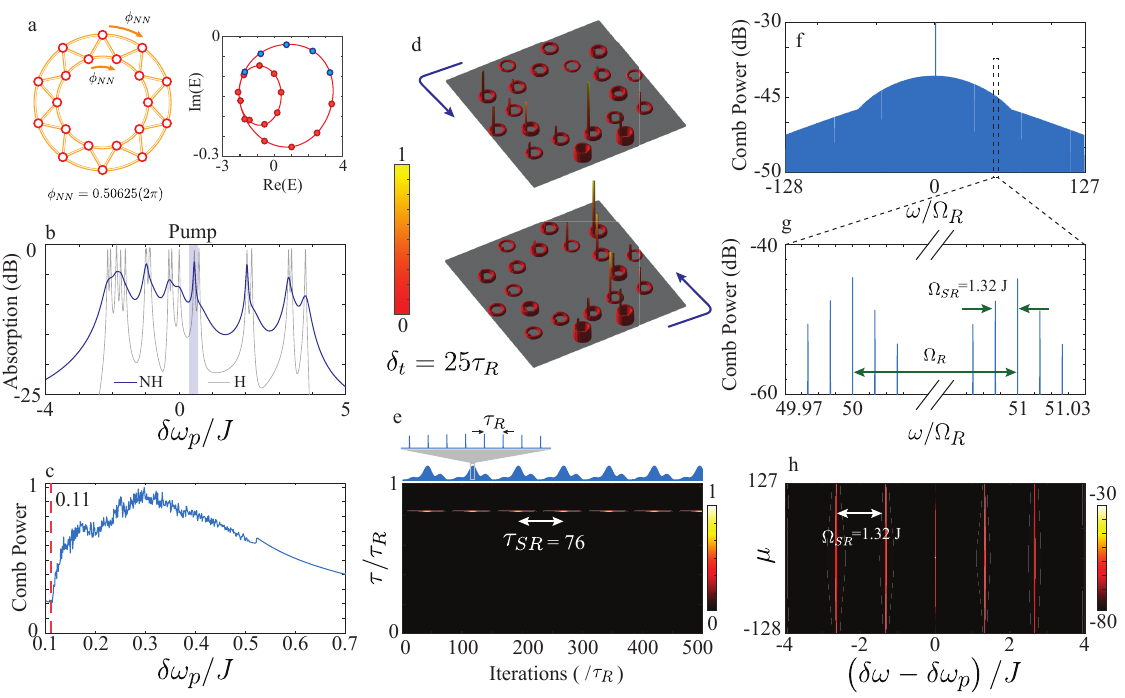}
 \caption{\textbf{Soliton combs in resonator arrays with nearest and next-nearest neighbor couplings, with phase $\phi_{NN} = 0.50625 (2\pi), ~J_{N} = J_{NN} = J, ~t^{NH}_{N} = 0.95, ~t^{NH}_{NN} = 0.9$}. \textbf{a}. Schematic of the array and the resultant double winding. The blue shaded dots on the winding curve indicate the low-dissipation supermodes that contribute to the nested comb formation. \textbf{b.} Linear absorption spectrum of the array for Hermitian and non-Hermitian lattices. The supermodes with low dissipation manifest as prominent peaks in the spectrum. \textbf{c.} Comb power as a function of pump frequency. The pump frequency where we observe nested solitons is indicated by the dashed line. \textbf{d.} Soliton intensity distribution in the array. As before, the soliton position within each ring hosting the super-soliton is the same.  With time, the nested soliton structure circulates the array in the CCW direction. \textbf{e.} Temporal output showing bursts of pulses, with each burst separated by $\tau_{SR} \simeq 76 \tau_{R}$. \textbf{f,g,h.} Generated comb spectrum showing the oscillation of five supermodes at each FSR, with FSR $\Omega_{SR} = 1.32 J$.}
\label{fig:3}
\end{figure*}
%%%%%%%%%%%%%%%%%%%%%%%%%%%%%%%%%%%%%%%%%%%%%%%%%%%%%%%%%%%%%%%%%%%%%%%%%%%%%

To generate nested combs, we pump the supermode in the center of the spectrum with the highest absorption. When we tune the pump frequency $\delta\omega_{p} = 0.11 J$ and set the normalized pump field $E_{in} = 0.11$, we observe the formation of a coherent comb characterized by a smooth intensity profile (Fig.3f). More importantly, within each FSR (Fig.3g), we now observe the oscillation of only five supermodes that correspond to the low-loss supermodes of the absorption spectrum. This contrasts the super-ring with only nearest-neighbor couplings, where we observed the oscillation of nine supermodes. Furthermore, the spacing between the oscillating supermodes $\Omega_{SR} \simeq 1.32 J$, which is much larger compared to Fig.2g,h where it was only about $0.59 J$. As before, we observe the complete cancellation of linear supermode dispersion against that of nonlinear dispersion, indicating the formation of coherent soliton combs (Fig.3h).

In the super-ring, we again observe the self-formation of phase-locked nested solitons, propagating around the lattice in the CCW direction (Fig.3d and Movie S2). These are qualitatively similar to those observed in the super-ring with only NN couplings (Fig.2d). The temporal output for this soliton comb also consists of bursts of pulses (Fig.3e), but the periodicity of the bursts $\tau_{SR} = 76 \tau_{R}$ is much smaller compared to Fig.2 where it was $168 \tau_{R}$. This decrease is consistent with the increase in the comb line spacing for NNN combs. Nevertheless, we also observe secondary bursts in the temporal output that have exactly the same periodicity as that of the primary bursts. This is because of the oscillation of a fewer number of supermodes that increases the spatio-temporal width of nested solitons, that is, they occupy more site rings and, therefore, generate broader temporal pulses.\\

%%%%%%%%%%%%%%%%%%%%%%%%%%%%%%%%%%%%%%%%%%%%%%%%%%%%%%%%%%%%%%%%%%%%%%%%%%%%%
% Figure 4
%%%%%%%%%%%%%%%%%%%%%%%%%%%%%%%%%%%%%%%%%%%%%%%%%%%%%%%%%%%%%%%%%%%%%%%%%%%%%
\begin{figure*}
 \centering
 \includegraphics[width=0.98\textwidth]{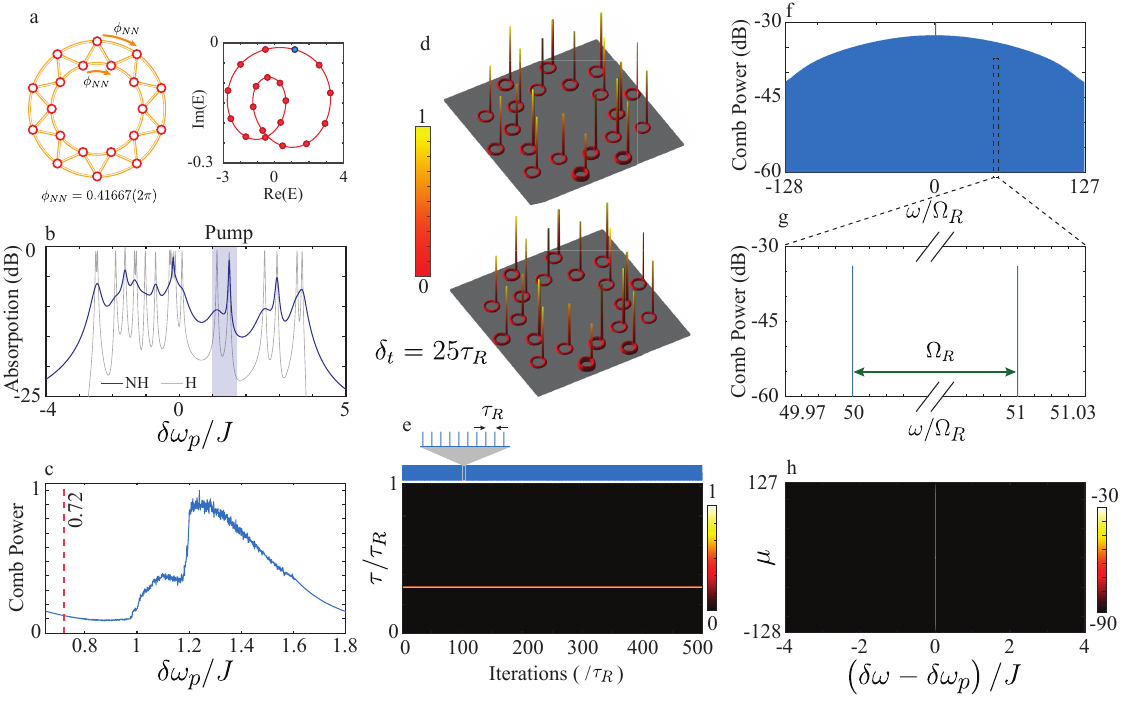}
 \caption{\textbf{Soliton combs in resonator arrays with nearest and next-nearest neighbor couplings, with phase phase $\phi_{NN} = 0.41667 (2\pi), ~J_{N} = J_{NN} = J, ~t^{NH}_{N} = 0.95, ~t^{NH}_{NN} = 0.9$}. \textbf{a.} Schematic of the array and the resultant double winding. The blue shaded dot on the winding curve indicates the pumped mode. \textbf{b.} Linear absorption spectrum of the array. The frequency spacings between neighboring supermodes with low dissipation are very different, which prevents the oscillation of multiple modes in the comb spectrum.  \textbf{c.} Comb power as a function of pump frequency. \textbf{d.} Soliton intensity distribution in the array. Each site ring of the array hosts exactly one soliton, and the soliton position within each ring is exactly the same. The soliton molecule structure does not show any evolution with time. \textbf{e.} Temporal output showing the generation of periodic light pulses separated by $\tau_{R}$. \textbf{f,g,h.} Generated comb spectrum where we observe the oscillation of only a single supermode at each FSR.}
\label{fig:4}
\end{figure*}
%%%%%%%%%%%%%%%%%%%%%%%%%%%%%%%%%%%%%%%%%%%%%%%%%%%%%%%%%%%%%%%%%%%%%%%%%%%%%

\noindent
\textbf{Tunability of the comb spectrum}\\
While post-fabrication tuning of the coupling rates between rings is challenging \cite{On2024}, the hopping phase can be tuned much more easily by, for example, incorporating thermal heaters on the link ring waveguides \cite{Mittal2016}. To show that tuning the hopping phase can lead to remarkable reconfigurability of the comb spectrum and the soliton state, in Fig.4, we show results for a super-ring with the same coupling parameters as in Fig.3, but we tune the hopping phase such that $\phi_{NN} = 0.41667 (2\pi)$. In this case, the non-Hermitian topological winding number stays the same, but the position of the inner winding loop rotates in the 2D complex energy plane (Fig.4a). The resulting absorption spectrum is shown in Fig.4b. Although the spectrum exhibits multiple peaks with low loss, our deliberate choice of $\phi_{NN}$ ensures that the frequency spacing between the consecutive peaks is very different.

When we pump the highlighted supermode, at $\delta\omega_{p} = 0.72 J$ and pump field $E_{in} = 0.11$, we now observe the oscillation of only a single supermode in the comb spectrum (Fig. 4f-h). This is in stark contrast with our previous demonstrations, where we observed nine and five supermodes in Fig.2 and Fig.3, respectively. This drastic reduction in the number of oscillating supermodes is due to our choice of supermode dispersion: no three supermode resonances are equally spaced in frequency to satisfy energy conservation for the FWM process. 

In the super-ring, we observe the formation of a new soliton molecule state where each site ring exhibits a single soliton (Fig.4d). As before, the solitons across different rings are phase-locked. This soliton-molecule state is stationary and does not show any evolution/circulation with time (Movie S3). This observation of a uniform spatial intensity distribution in the super-ring is consistent with the oscillation of a single supermode in the comb spectrum. The temporal output of this super-ring shows periodic pulses with a repetition rate of $\tau_{R}$ (Fig.4e). There are no bursts of pulses as were observed with nested solitons. We note that this output temporal and frequency spectrum of this lattice is exactly the same as that of a single ring resonator: comb lines separated by $\Omega_{R}$ and temporal pulses separated by $\tau_{R} = \frac{1}{\Omega_{R}}$. However, this spectrum is generated by a set of 20 self-synchronized ring resonators. 

%%%%%%%%%%%%%%%%%%%%%%%%%%%%%%%%%%%%%%%%%%%%%%%%%%%%%%%%%%%%%%%%%%%%%%%%%%%%%
% Discussion and Conclusion
%%%%%%%%%%%%%%%%%%%%%%%%%%%%%%%%%%%%%%%%%%%%%%%%%%%%%%%%%%%%%%%%%%%%%%%%%%%%%
\section*{Discussion}
To summarize, we have demonstrated the use of dissipative couplings to engineer the generation of soliton combs in 1D resonator arrays with periodic boundary conditions. Although we have explored only a very small subset of parameter choices, this system offers many degrees of freedom in the form of relative strength of nearest and next-nearest couplings $J_{N}/J_{NN}$, relative phase $\phi_{NN}$ and $\phi_{N}$, and even the strength of dissipation $\delta_{N}$ and $\delta_{NN}$. The system could be designed using another choice of fixed parameters $\left(J_{N}, J_{NN}, \delta_{N}, \delta_{NN} \right)$ such that tuning hopping phases could lead to even more drastic agile reconfigurability of the comb spectrum. This operating regime could also be set to suit a particular comb application. As examples, in the Supplementary Materials, we show results for other parameter choices where we observe the formation of a nested comb with three and seven oscillating supermodes in each FSR. 

Our approach is very general and can be easily extended to generate new soliton comb states in 2D coupled ring resonator arrays where, for example, first-order or higher-order non-Hermitian physics could lead to topologically nontrivial states of light \cite{Hafezi2011, Hafezi2013, Mittal2019, Mittal2019b, Afzal2020, Ozawa2024, Luo2019, Edvardsson2019, Ozawa2019}. Although we have focused on Kerr combs, non-Hermitian engineering can also be applied to engineer, for example,  electro-optic or optomechanical combs \cite{Zhang2019, Rueda2019, Zhang2021, Hussein2024}. Therefore, our results pave the way to explore the rich synergy between nonlinear dynamics and the formation of frequency combs, synthetic dimensions, and the physics of driven dissipative non-Hermitian systems.\\

%%%%%%%%%%%%%%%%%%%%%%%%%%%%%%%%%%%%%%%%%%%%%%%%%%%%%%%%%%%%%%%%%%%%%%%%%%%%%
% Discussion and Conclusion
%%%%%%%%%%%%%%%%%%%%%%%%%%%%%%%%%%%%%%%%%%%%%%%%%%%%%%%%%%%%%%%%%%%%%%%%%%%%%
\noindent
\textbf{Acknowledgements}
This research was supported by startup and TIER 1 grants from Northeastern University. \\

\noindent
\textbf{Author Contributions}
S.M. conceived the idea and developed the simulation framework. S.D.H. performed the numerical simulations. Both authors contributed to the analysis of the data. S.D.H. and S.M. prepared the figures. S.M. wrote the manuscript with input from S.D.H.. S.M. supervised the project.\\

%%%%%%%%%%%%%%%%%%%%%%%%%%%%%%%%%%%%%%%%%%%%%%%%%%%%%%%%%%%%%%%%%%%%%%%%%%%%%
% bibliography
%%%%%%%%%%%%%%%%%%%%%%%%%%%%%%%%%%%%%%%%%%%%%%%%%%%%%%%%%%%%%%%%%%%%%%%%%%%%%

%\bibliographystyle{NatureMag}
%\bibliography{FC_Biblio, Topo_Biblio, NH_Biblio}

\bibliography{Main_NH_Comb_arXiv.bbl}

\providecommand{\noopsort}[1]{}\providecommand{\singleletter}[1]{#1}%\providecommand{\noopsort}[1]{}\providecommand{\singleletter}[1]{#1}%\providecommand{\noopsort}[1]{}\providecommand{\singleletter}[1]{#1}%
\begin{thebibliography}{10}
\expandafter\ifx\csname url\endcsname\relax
  \def\url#1{\texttt{#1}}\fi
\expandafter\ifx\csname urlprefix\endcsname\relax\def\urlprefix{URL }\fi
\providecommand{\bibinfo}[2]{#2}
\providecommand{\eprint}[2][]{\url{#2}}

\bibitem{Cundiff2003}
\bibinfo{author}{Cundiff, S.~T.} \& \bibinfo{author}{Ye, J.}
\newblock \bibinfo{title}{Colloquium: Femtosecond optical frequency combs}.
\newblock \emph{\bibinfo{journal}{Rev. Mod. Phys.}}
  \textbf{\bibinfo{volume}{75}}, \bibinfo{pages}{325--342}
  (\bibinfo{year}{2003}).

\bibitem{Kippenberg2011}
\bibinfo{author}{Kippenberg, T.~J.}, \bibinfo{author}{Holzwarth, R.} \&
  \bibinfo{author}{Diddams, S.~A.}
\newblock \bibinfo{title}{Microresonator-based optical frequency combs}.
\newblock \emph{\bibinfo{journal}{Science}} \textbf{\bibinfo{volume}{332}},
  \bibinfo{pages}{555--559} (\bibinfo{year}{2011}).

\bibitem{Kippenberg2018}
\bibinfo{author}{Kippenberg, T.~J.}, \bibinfo{author}{Gaeta, A.~L.},
  \bibinfo{author}{Lipson, M.} \& \bibinfo{author}{Gorodetsky, M.~L.}
\newblock \bibinfo{title}{Dissipative kerr solitons in optical
  microresonators}.
\newblock \emph{\bibinfo{journal}{Science}} \textbf{\bibinfo{volume}{361}}
  (\bibinfo{year}{2018}).

\bibitem{Pasquazi2018}
\bibinfo{author}{Pasquazi, A.} \emph{et~al.}
\newblock \bibinfo{title}{Micro-combs: A novel generation of optical sources}.
\newblock \emph{\bibinfo{journal}{Phys. Rep.}} \textbf{\bibinfo{volume}{729}},
  \bibinfo{pages}{1--81} (\bibinfo{year}{2018}).

\bibitem{Gaeta2019}
\bibinfo{author}{Gaeta, A.~L.}, \bibinfo{author}{Lipson, M.} \&
  \bibinfo{author}{Kippenberg, T.~J.}
\newblock \bibinfo{title}{Photonic-chip-based frequency combs}.
\newblock \emph{\bibinfo{journal}{Nat. Photonics}}
  \textbf{\bibinfo{volume}{13}}, \bibinfo{pages}{158--169}
  (\bibinfo{year}{2019}).

\bibitem{Anderson2022}
\bibinfo{author}{Anderson, M.~H.} \emph{et~al.}
\newblock \bibinfo{title}{Zero dispersion kerr solitons in optical
  microresonators}.
\newblock \emph{\bibinfo{journal}{Nat. Commun.}} \textbf{\bibinfo{volume}{13}},
  \bibinfo{pages}{4764} (\bibinfo{year}{2022}).

\bibitem{Yu2021}
\bibinfo{author}{Yu, S.-P.} \emph{et~al.}
\newblock \bibinfo{title}{Spontaneous pulse formation in edgeless photonic
  crystal resonators}.
\newblock \emph{\bibinfo{journal}{Nat. Photonics}}
  \textbf{\bibinfo{volume}{15}}, \bibinfo{pages}{461--467}
  (\bibinfo{year}{2021}).

\bibitem{Yu2022}
\bibinfo{author}{Yu, S.-P.}, \bibinfo{author}{Lucas, E.},
  \bibinfo{author}{Zang, J.} \& \bibinfo{author}{Papp, S.~B.}
\newblock \bibinfo{title}{A continuum of bright and dark-pulse states in a
  photonic-crystal resonator}.
\newblock \emph{\bibinfo{journal}{Nat. Commun.}} \textbf{\bibinfo{volume}{13}},
  \bibinfo{pages}{3134} (\bibinfo{year}{2022}).

\bibitem{Lucas2023}
\bibinfo{author}{Lucas, E.}, \bibinfo{author}{Yu, S.-P.},
  \bibinfo{author}{Briles, T.~C.}, \bibinfo{author}{Carlson, D.~R.} \&
  \bibinfo{author}{Papp, S.~B.}
\newblock \bibinfo{title}{Tailoring microcombs with inverse-designed,
  meta-dispersion microresonators}.
\newblock \emph{\bibinfo{journal}{Nat. Photonics}}
  \textbf{\bibinfo{volume}{17}}, \bibinfo{pages}{943--950}
  (\bibinfo{year}{2023}).

\bibitem{Yang2023}
\bibinfo{author}{Yang, J.}, \bibinfo{author}{Guidry, M.~A.},
  \bibinfo{author}{Lukin, D.~M.}, \bibinfo{author}{Yang, K.} \&
  \bibinfo{author}{Vučković, J.}
\newblock \bibinfo{title}{Inverse-designed silicon carbide quantum and
  nonlinear photonics}.
\newblock \emph{\bibinfo{journal}{Light Sci Appl}}
  \textbf{\bibinfo{volume}{12}}, \bibinfo{pages}{201} (\bibinfo{year}{2023}).

\bibitem{Miller2015}
\bibinfo{author}{Miller, S.~A.} \emph{et~al.}
\newblock \bibinfo{title}{Tunable frequency combs based on dual microring
  resonators}.
\newblock \emph{\bibinfo{journal}{Opt. Express, OE}}
  \textbf{\bibinfo{volume}{23}}, \bibinfo{pages}{21527--21540}
  (\bibinfo{year}{2015}).

\bibitem{Kim2017}
\bibinfo{author}{Kim, S.} \emph{et~al.}
\newblock \bibinfo{title}{Dispersion engineering and frequency comb generation
  in thin silicon nitride concentric microresonators}.
\newblock \emph{\bibinfo{journal}{Nat. Commun.}} \textbf{\bibinfo{volume}{8}},
  \bibinfo{pages}{372} (\bibinfo{year}{2017}).

\bibitem{Jang2018}
\bibinfo{author}{Jang, J.~K.} \emph{et~al.}
\newblock \bibinfo{title}{Synchronization of coupled optical microresonators}.
\newblock \emph{\bibinfo{journal}{Nat. Photonics}}
  \textbf{\bibinfo{volume}{12}}, \bibinfo{pages}{688--693}
  (\bibinfo{year}{2018}).

\bibitem{Vasco2019}
\bibinfo{author}{Vasco, J.} \& \bibinfo{author}{Savona, V.}
\newblock \bibinfo{title}{Slow-light frequency combs and dissipative kerr
  solitons in coupled-cavity waveguides}.
\newblock \emph{\bibinfo{journal}{Phys. Rev. Applied}}
  \textbf{\bibinfo{volume}{12}}, \bibinfo{pages}{064065}
  (\bibinfo{year}{2019}).

\bibitem{Helgason2021}
\bibinfo{author}{Helgason, {\'O}.~B.} \emph{et~al.}
\newblock \bibinfo{title}{Dissipative solitons in photonic molecules}.
\newblock \emph{\bibinfo{journal}{Nat. Photonics}}  (\bibinfo{year}{2021}).

\bibitem{Tikan2021}
\bibinfo{author}{Tikan, A.} \emph{et~al.}
\newblock \bibinfo{title}{Emergent nonlinear phenomena in a driven dissipative
  photonic dimer}.
\newblock \emph{\bibinfo{journal}{Nat. Phys.}} \textbf{\bibinfo{volume}{17}},
  \bibinfo{pages}{604--610} (\bibinfo{year}{2021}).

\bibitem{Yuan2023}
\bibinfo{author}{Yuan, Z.} \emph{et~al.}
\newblock \bibinfo{title}{Soliton pulse pairs at multiple colours in normal
  dispersion microresonators}.
\newblock \emph{\bibinfo{journal}{Nat. Photonics}}
  \textbf{\bibinfo{volume}{17}}, \bibinfo{pages}{977--983}
  (\bibinfo{year}{2023}).

\bibitem{Helgason2023}
\bibinfo{author}{Helgason, O.~B.} \emph{et~al.}
\newblock \bibinfo{title}{Surpassing the nonlinear conversion efficiency of
  soliton microcombs}.
\newblock \emph{\bibinfo{journal}{Nat. Photonics}}
  \textbf{\bibinfo{volume}{17}}, \bibinfo{pages}{992--999}
  (\bibinfo{year}{2023}).

\bibitem{Mittal2021b}
\bibinfo{author}{Mittal, S.}, \bibinfo{author}{Moille, G.},
  \bibinfo{author}{Srinivasan, K.}, \bibinfo{author}{Chembo, Y.~K.} \&
  \bibinfo{author}{Hafezi, M.}
\newblock \bibinfo{title}{Topological frequency combs and nested temporal
  solitons}.
\newblock \emph{\bibinfo{journal}{Nat. Phys.}} \textbf{\bibinfo{volume}{17}},
  \bibinfo{pages}{1169--1176} (\bibinfo{year}{2021}).

\bibitem{Flower2024}
\bibinfo{author}{Flower, C.~J.} \emph{et~al.}
\newblock \bibinfo{title}{Observation of topological frequency combs}.
\newblock \emph{\bibinfo{journal}{Science}} \textbf{\bibinfo{volume}{384}},
  \bibinfo{pages}{1356--1361} (\bibinfo{year}{2024}).

\bibitem{Tusnin2023}
\bibinfo{author}{Tusnin, A.}, \bibinfo{author}{Tikan, A.},
  \bibinfo{author}{Komagata, K.} \& \bibinfo{author}{Kippenberg, T.~J.}
\newblock \bibinfo{title}{Nonlinear dynamics and kerr frequency comb formation
  in lattices of coupled microresonators}.
\newblock \emph{\bibinfo{journal}{Communications Physics}}
  \textbf{\bibinfo{volume}{6}}, \bibinfo{pages}{1--10} (\bibinfo{year}{2023}).

\bibitem{Hashemi2024}
\bibinfo{author}{Hashemi, S.~D.} \& \bibinfo{author}{Mittal, S.}
\newblock \bibinfo{title}{Floquet topological dissipative kerr solitons and
  incommensurate frequency combs}.
\newblock \emph{\bibinfo{journal}{Nat. Commun.}} \textbf{\bibinfo{volume}{15}},
  \bibinfo{pages}{1--9} (\bibinfo{year}{2024}).

\bibitem{Gong2018}
\bibinfo{author}{Gong, Z.} \emph{et~al.}
\newblock \bibinfo{title}{Topological phases of non-hermitian systems}.
\newblock \emph{\bibinfo{journal}{Phys. Rev. X}} \textbf{\bibinfo{volume}{8}},
  \bibinfo{pages}{031079} (\bibinfo{year}{2018}).

\bibitem{Ashida2020}
\bibinfo{author}{Ashida, Y.}, \bibinfo{author}{Gong, Z.} \&
  \bibinfo{author}{Ueda, M.}
\newblock \bibinfo{title}{Non-hermitian physics}.
\newblock \emph{\bibinfo{journal}{Adv. Phys.}} \textbf{\bibinfo{volume}{69}},
  \bibinfo{pages}{249--435} (\bibinfo{year}{2020}).

\bibitem{Okuma2020}
\bibinfo{author}{Okuma, N.}, \bibinfo{author}{Kawabata, K.},
  \bibinfo{author}{Shiozaki, K.} \& \bibinfo{author}{Sato, M.}
\newblock \bibinfo{title}{Topological origin of non-hermitian skin effects}.
\newblock \emph{\bibinfo{journal}{Phys. Rev. Lett.}}
  \textbf{\bibinfo{volume}{124}}, \bibinfo{pages}{086801}
  (\bibinfo{year}{2020}).

\bibitem{Kawabata2019}
\bibinfo{author}{Kawabata, K.}, \bibinfo{author}{Shiozaki, K.},
  \bibinfo{author}{Ueda, M.} \& \bibinfo{author}{Sato, M.}
\newblock \bibinfo{title}{Symmetry and topology in non-hermitian physics}.
\newblock \emph{\bibinfo{journal}{Phys. Rev. X}} \textbf{\bibinfo{volume}{9}},
  \bibinfo{pages}{041015} (\bibinfo{year}{2019}).

\bibitem{Okuma2023}
\bibinfo{author}{Okuma, N.} \& \bibinfo{author}{Sato, M.}
\newblock \bibinfo{title}{Non-hermitian topological phenomena: A review}.
\newblock \emph{\bibinfo{journal}{Annu. Rev. Condens. Matter Phys.}}
  \textbf{\bibinfo{volume}{14}}, \bibinfo{pages}{83--107}
  (\bibinfo{year}{2023}).

\bibitem{El-Ganainy2018}
\bibinfo{author}{El-Ganainy, R.} \emph{et~al.}
\newblock \bibinfo{title}{Non-hermitian physics and {PT} symmetry}.
\newblock \emph{\bibinfo{journal}{Nat. Phys.}} \textbf{\bibinfo{volume}{14}},
  \bibinfo{pages}{11--19} (\bibinfo{year}{2018}).

\bibitem{Miri2019}
\bibinfo{author}{Miri, M.-A.} \& \bibinfo{author}{Alù, A.}
\newblock \bibinfo{title}{Exceptional points in optics and photonics}.
\newblock \emph{\bibinfo{journal}{Science}} \textbf{\bibinfo{volume}{363}},
  \bibinfo{pages}{eaar7709} (\bibinfo{year}{2019}).

\bibitem{Nasari2023}
\bibinfo{author}{Nasari, H.}, \bibinfo{author}{Pyrialakos, G.~G.},
  \bibinfo{author}{Christodoulides, D.~N.} \& \bibinfo{author}{Khajavikhan, M.}
\newblock \bibinfo{title}{Non-hermitian topological photonics}.
\newblock \emph{\bibinfo{journal}{Opt. Mater. Express}}
  \textbf{\bibinfo{volume}{13}}, \bibinfo{pages}{870} (\bibinfo{year}{2023}).

\bibitem{Li2023}
\bibinfo{author}{Li, Z.} \emph{et~al.}
\newblock \bibinfo{title}{Synergetic positivity of loss and noise in nonlinear
  non-hermitian resonators}.
\newblock \emph{\bibinfo{journal}{Sci. Adv.}} \textbf{\bibinfo{volume}{9}},
  \bibinfo{pages}{eadi0562} (\bibinfo{year}{2023}).

\bibitem{Wang2023}
\bibinfo{author}{Wang, Q.} \& \bibinfo{author}{Chong, Y.~D.}
\newblock \bibinfo{title}{Non-hermitian photonic lattices: tutorial}.
\newblock \emph{\bibinfo{journal}{J. Opt. Soc. Am. B}}
  \textbf{\bibinfo{volume}{40}}, \bibinfo{pages}{1443} (\bibinfo{year}{2023}).

\bibitem{Fang2017}
\bibinfo{author}{Fang, K.} \emph{et~al.}
\newblock \bibinfo{title}{Generalized non-reciprocity in an optomechanical
  circuit via synthetic magnetism and reservoir engineering}.
\newblock \emph{\bibinfo{journal}{Nat. Phys.}} \textbf{\bibinfo{volume}{13}},
  \bibinfo{pages}{465--471} (\bibinfo{year}{2017}).

\bibitem{Metelmann2015}
\bibinfo{author}{Metelmann, A.} \& \bibinfo{author}{Clerk, A.~A.}
\newblock \bibinfo{title}{Nonreciprocal photon transmission and amplification
  via reservoir engineering}.
\newblock \emph{\bibinfo{journal}{Phys. Rev. X}} \textbf{\bibinfo{volume}{5}},
  \bibinfo{pages}{021025} (\bibinfo{year}{2015}).

\bibitem{Weidemann2020}
\bibinfo{author}{Weidemann, S.} \emph{et~al.}
\newblock \bibinfo{title}{Topological funneling of light}.
\newblock \emph{\bibinfo{journal}{Science}} \textbf{\bibinfo{volume}{368}},
  \bibinfo{pages}{311--314} (\bibinfo{year}{2020}).

\bibitem{Leefmans2022}
\bibinfo{author}{Leefmans, C.} \emph{et~al.}
\newblock \bibinfo{title}{Topological dissipation in a time-multiplexed
  photonic resonator network}.
\newblock \emph{\bibinfo{journal}{Nat. Phys.}} \textbf{\bibinfo{volume}{18}},
  \bibinfo{pages}{442--449} (\bibinfo{year}{2022}).

\bibitem{Ding2022}
\bibinfo{author}{Ding, K.}, \bibinfo{author}{Fang, C.} \& \bibinfo{author}{Ma,
  G.}
\newblock \bibinfo{title}{Non-hermitian topology and exceptional-point
  geometries}.
\newblock \emph{\bibinfo{journal}{Nature Reviews Physics}}
  \textbf{\bibinfo{volume}{4}}, \bibinfo{pages}{745--760}
  (\bibinfo{year}{2022}).

\bibitem{Wang2021}
\bibinfo{author}{Wang, K.} \emph{et~al.}
\newblock \bibinfo{title}{Generating arbitrary topological windings of a
  non-hermitian band}.
\newblock \emph{\bibinfo{journal}{Science}} \textbf{\bibinfo{volume}{371}},
  \bibinfo{pages}{1240--1245} (\bibinfo{year}{2021}).

\bibitem{Wang2021b}
\bibinfo{author}{Wang, K.}, \bibinfo{author}{Dutt, A.},
  \bibinfo{author}{Wojcik, C.~C.} \& \bibinfo{author}{Fan, S.}
\newblock \bibinfo{title}{Topological complex-energy braiding of non-hermitian
  bands}.
\newblock \emph{\bibinfo{journal}{Nature}} \textbf{\bibinfo{volume}{598}},
  \bibinfo{pages}{59--64} (\bibinfo{year}{2021}).

\bibitem{Chembo2016}
\bibinfo{author}{Chembo, Y.~K.}
\newblock \bibinfo{title}{Kerr optical frequency combs: theory, applications
  and perspectives}.
\newblock \emph{\bibinfo{journal}{Nanophotonics}} \textbf{\bibinfo{volume}{5}},
  \bibinfo{pages}{7957} (\bibinfo{year}{2016}).

\bibitem{Pernet2022}
\bibinfo{author}{Pernet, N.} \emph{et~al.}
\newblock \bibinfo{title}{Gap solitons in a one-dimensional driven-dissipative
  topological lattice}.
\newblock \emph{\bibinfo{journal}{Nat. Phys.}} \textbf{\bibinfo{volume}{18}},
  \bibinfo{pages}{678--684} (\bibinfo{year}{2022}).

\bibitem{Xia2021}
\bibinfo{author}{Xia, S.} \emph{et~al.}
\newblock \bibinfo{title}{Nonlinear tuning of {PT} symmetry and non-hermitian
  topological states}.
\newblock \emph{\bibinfo{journal}{Science}} \textbf{\bibinfo{volume}{372}},
  \bibinfo{pages}{72--76} (\bibinfo{year}{2021}).

\bibitem{Dai2023}
\bibinfo{author}{Dai, T.} \emph{et~al.}
\newblock \bibinfo{title}{Non-hermitian topological phase transitions
  controlled by nonlinearity}.
\newblock \emph{\bibinfo{journal}{Nat. Phys.}} \textbf{\bibinfo{volume}{20}},
  \bibinfo{pages}{101--108} (\bibinfo{year}{2023}).

\bibitem{Reisenbauer2024}
\bibinfo{author}{Reisenbauer, M.} \emph{et~al.}
\newblock \bibinfo{title}{Non-hermitian dynamics and non-reciprocity of
  optically coupled nanoparticles}.
\newblock \emph{\bibinfo{journal}{Nat. Phys.}} \textbf{\bibinfo{volume}{20}},
  \bibinfo{pages}{1629--1635} (\bibinfo{year}{2024}).

\bibitem{Liu2022}
\bibinfo{author}{Liu, Y. G.~N.} \emph{et~al.}
\newblock \bibinfo{title}{Complex skin modes in non-hermitian coupled laser
  arrays}.
\newblock \emph{\bibinfo{journal}{Light Sci Appl}}
  \textbf{\bibinfo{volume}{11}}, \bibinfo{pages}{336} (\bibinfo{year}{2022}).

\bibitem{Leefmans2024}
\bibinfo{author}{Leefmans, C.~R.} \emph{et~al.}
\newblock \bibinfo{title}{Topological temporally mode-locked laser}.
\newblock \emph{\bibinfo{journal}{Nat. Phys.}} \textbf{\bibinfo{volume}{20}},
  \bibinfo{pages}{852--858} (\bibinfo{year}{2024}).

\bibitem{Hafezi2011}
\bibinfo{author}{Hafezi, M.}, \bibinfo{author}{Demler, E.~A.},
  \bibinfo{author}{Lukin, M.~D.} \& \bibinfo{author}{Taylor, J.~M.}
\newblock \bibinfo{title}{Robust optical delay lines with topological
  protection}.
\newblock \emph{\bibinfo{journal}{Nature Physics}}
  \textbf{\bibinfo{volume}{7}}, \bibinfo{pages}{907} (\bibinfo{year}{2011}).

\bibitem{Hafezi2013}
\bibinfo{author}{Hafezi, M.}, \bibinfo{author}{Mittal, S.},
  \bibinfo{author}{Fan, J.}, \bibinfo{author}{Migdall, A.} \&
  \bibinfo{author}{Taylor, J.}
\newblock \bibinfo{title}{Imaging topological edge states in silicon
  photonics}.
\newblock \emph{\bibinfo{journal}{Nature Photonics}}
  \textbf{\bibinfo{volume}{7}}, \bibinfo{pages}{1001} (\bibinfo{year}{2013}).

\bibitem{Ikeda1979}
\bibinfo{author}{Ikeda, K.}
\newblock \bibinfo{title}{Multiple-valued stationary state and its instability
  of the transmitted light by a ring cavity system}.
\newblock \emph{\bibinfo{journal}{Opt. Commun.}} \textbf{\bibinfo{volume}{30}},
  \bibinfo{pages}{257--261} (\bibinfo{year}{1979}).

\bibitem{Hansson2015}
\bibinfo{author}{Hansson, T.} \& \bibinfo{author}{Wabnitz, S.}
\newblock \bibinfo{title}{Frequency comb generation beyond the
  {Lugiato--Lefever} equation: multi-stability and super cavity solitons}.
\newblock \emph{\bibinfo{journal}{J. Opt. Soc. Am. B, JOSAB}}
  \textbf{\bibinfo{volume}{32}}, \bibinfo{pages}{1259--1266}
  (\bibinfo{year}{2015}).

\bibitem{Hansson2016}
\bibinfo{author}{Hansson, T.} \& \bibinfo{author}{Wabnitz, S.}
\newblock \bibinfo{title}{Dynamics of microresonator frequency comb generation:
  models and stability}.
\newblock \emph{\bibinfo{journal}{Nanophotonics}} \textbf{\bibinfo{volume}{5}},
  \bibinfo{pages}{231--243} (\bibinfo{year}{2016}).

\bibitem{Little1997}
\bibinfo{author}{{Little}, B.~E.}, \bibinfo{author}{{Chu}, S.~T.},
  \bibinfo{author}{{Haus}, H.~A.}, \bibinfo{author}{{Foresi}, J.} \&
  \bibinfo{author}{{Laine}, J.~.}
\newblock \bibinfo{title}{Microring resonator channel dropping filters}.
\newblock \emph{\bibinfo{journal}{Journal of Lightwave Technology}}
  \textbf{\bibinfo{volume}{15}}, \bibinfo{pages}{998--1005}
  (\bibinfo{year}{1997}).

\bibitem{Chembo2013}
\bibinfo{author}{Chembo, Y.~K.} \& \bibinfo{author}{Menyuk, C.~R.}
\newblock \bibinfo{title}{Spatiotemporal lugiato-lefever formalism for
  kerr-comb generation in whispering-gallery-mode resonators}.
\newblock \emph{\bibinfo{journal}{Phys. Rev. A}} \textbf{\bibinfo{volume}{87}},
  \bibinfo{pages}{053852} (\bibinfo{year}{2013}).

\bibitem{Mittal2021}
\bibinfo{author}{Mittal, S.}, \bibinfo{author}{Orre, V.~V.},
  \bibinfo{author}{Goldschmidt, E.~A.} \& \bibinfo{author}{Hafezi, M.}
\newblock \bibinfo{title}{Tunable quantum interference using a topological
  source of indistinguishable photon pairs}.
\newblock \emph{\bibinfo{journal}{Nat. Photonics}}
  \textbf{\bibinfo{volume}{15}}, \bibinfo{pages}{542--548}
  (\bibinfo{year}{2021-05-10}).

\bibitem{On2024}
\bibinfo{author}{On, M.~B.} \emph{et~al.}
\newblock \bibinfo{title}{Programmable integrated photonics for topological
  hamiltonians}.
\newblock \emph{\bibinfo{journal}{Nat. Commun.}} \textbf{\bibinfo{volume}{15}},
  \bibinfo{pages}{629} (\bibinfo{year}{2024}).

\bibitem{Mittal2016}
\bibinfo{author}{Mittal, S.}, \bibinfo{author}{Ganeshan, S.},
  \bibinfo{author}{Fan, J.}, \bibinfo{author}{Vaezi, A.} \&
  \bibinfo{author}{Hafezi, M.}
\newblock \bibinfo{title}{{Measurement of topological invariants in a 2D
  photonic system}}.
\newblock \emph{\bibinfo{journal}{Nat. Photon.}} \textbf{\bibinfo{volume}{10}},
  \bibinfo{pages}{180--183} (\bibinfo{year}{2016}).

\bibitem{Mittal2019}
\bibinfo{author}{Mittal, S.}, \bibinfo{author}{Orre, V.~V.},
  \bibinfo{author}{Leykam, D.}, \bibinfo{author}{Chong, Y.~D.} \&
  \bibinfo{author}{Hafezi, M.}
\newblock \bibinfo{title}{Photonic anomalous quantum hall effect}.
\newblock \emph{\bibinfo{journal}{Phys. Rev. Lett.}}
  \textbf{\bibinfo{volume}{123}}, \bibinfo{pages}{043201}
  (\bibinfo{year}{2019}).

\bibitem{Mittal2019b}
\bibinfo{author}{Mittal, S.} \emph{et~al.}
\newblock \bibinfo{title}{Photonic quadrupole topological phases}.
\newblock \emph{\bibinfo{journal}{Nat. Photonics}}
  \textbf{\bibinfo{volume}{13}}, \bibinfo{pages}{692--696}
  (\bibinfo{year}{2019}).

\bibitem{Afzal2020}
\bibinfo{author}{Afzal, S.}, \bibinfo{author}{Zimmerling, T.~J.},
  \bibinfo{author}{Ren, Y.}, \bibinfo{author}{Perron, D.} \&
  \bibinfo{author}{Van, V.}
\newblock \bibinfo{title}{Realization of anomalous floquet insulators in
  strongly coupled nanophotonic lattices}.
\newblock \emph{\bibinfo{journal}{Phys. Rev. Lett.}}
  \textbf{\bibinfo{volume}{124}}, \bibinfo{pages}{253601}
  (\bibinfo{year}{2020}).

\bibitem{Ozawa2024}
\bibinfo{author}{Ozawa, T.} \& \bibinfo{author}{Hayata, T.}
\newblock \bibinfo{title}{Two-dimensional lattice with an imaginary magnetic
  field}.
\newblock \emph{\bibinfo{journal}{Phys. Rev. B.}}
  \textbf{\bibinfo{volume}{109}}, \bibinfo{pages}{085113}
  (\bibinfo{year}{2024}).

\bibitem{Luo2019}
\bibinfo{author}{Luo, X.-W.} \& \bibinfo{author}{Zhang, C.}
\newblock \bibinfo{title}{Higher-order topological corner states induced by
  gain and loss}.
\newblock \emph{\bibinfo{journal}{Phys. Rev. Lett.}}
  \textbf{\bibinfo{volume}{123}}, \bibinfo{pages}{073601}
  (\bibinfo{year}{2019}).

\bibitem{Edvardsson2019}
\bibinfo{author}{Edvardsson, E.}, \bibinfo{author}{Kunst, F.~K.} \&
  \bibinfo{author}{Bergholtz, E.~J.}
\newblock \bibinfo{title}{Non-hermitian extensions of higher-order topological
  phases and their biorthogonal bulk-boundary correspondence}.
\newblock \emph{\bibinfo{journal}{Phys. Rev. B}} \textbf{\bibinfo{volume}{99}},
  \bibinfo{pages}{081302} (\bibinfo{year}{2019}).

\bibitem{Ozawa2019}
\bibinfo{author}{Ozawa, T.} \emph{et~al.}
\newblock \bibinfo{title}{Topological photonics}.
\newblock \emph{\bibinfo{journal}{Rev. Mod. Phys.}}
  \textbf{\bibinfo{volume}{91}}, \bibinfo{pages}{015006}
  (\bibinfo{year}{2019}).

\bibitem{Zhang2019}
\bibinfo{author}{Zhang, M.} \emph{et~al.}
\newblock \bibinfo{title}{Broadband electro-optic frequency comb generation in
  a lithium niobate microring resonator}.
\newblock \emph{\bibinfo{journal}{Nature}} \textbf{\bibinfo{volume}{568}},
  \bibinfo{pages}{373--377} (\bibinfo{year}{2019}).

\bibitem{Rueda2019}
\bibinfo{author}{Rueda, A.}, \bibinfo{author}{Sedlmeir, F.},
  \bibinfo{author}{Kumari, M.}, \bibinfo{author}{Leuchs, G.} \&
  \bibinfo{author}{Schwefel, H. G.~L.}
\newblock \bibinfo{title}{Resonant electro-optic frequency comb}.
\newblock \emph{\bibinfo{journal}{Nature}} \textbf{\bibinfo{volume}{568}},
  \bibinfo{pages}{378--381} (\bibinfo{year}{2019}).

\bibitem{Zhang2021}
\bibinfo{author}{Zhang, J.} \emph{et~al.}
\newblock \bibinfo{title}{Optomechanical dissipative solitons}.
\newblock \emph{\bibinfo{journal}{Nature}} \textbf{\bibinfo{volume}{600}},
  \bibinfo{pages}{75--80} (\bibinfo{year}{2021}).

\bibitem{Hussein2024}
\bibinfo{author}{Hussein, H. M.~E.}, \bibinfo{author}{Kim, S.},
  \bibinfo{author}{Rinaldi, M.}, \bibinfo{author}{Alù, A.} \&
  \bibinfo{author}{Cassella, C.}
\newblock \bibinfo{title}{Passive frequency comb generation at radiofrequency
  for ranging applications}.
\newblock \emph{\bibinfo{journal}{Nat. Commun.}} \textbf{\bibinfo{volume}{15}},
  \bibinfo{pages}{2844} (\bibinfo{year}{2024}).

\bibitem{NLOptics_Boyd}
\bibinfo{author}{Boyd, R.~W.}
\newblock \emph{\bibinfo{title}{Nonlinear Optics, Third Edition}}
  (\bibinfo{publisher}{Academic Press, Inc.}, \bibinfo{address}{USA},
  \bibinfo{year}{2008}), \bibinfo{edition}{3rd} edn.

\bibitem{Pfeiffer2017}
\bibinfo{author}{Pfeiffer, M. H.~P.} \emph{et~al.}
\newblock \bibinfo{title}{Octave-spanning dissipative kerr soliton frequency
  combs in $\text{Si}_{3}\text{N}_{4}$ microresonators}.
\newblock \emph{\bibinfo{journal}{Optica}} \textbf{\bibinfo{volume}{4}},
  \bibinfo{pages}{684} (\bibinfo{year}{2017}).

\end{thebibliography}


\providecommand{\noopsort}[1]{}\providecommand{\singleletter}[1]{#1}%\providecommand{\noopsort}[1]{}\providecommand{\singleletter}[1]{#1}%\providecommand{\noopsort}[1]{}\providecommand{\singleletter}[1]{#1}%
\begin{thebibliography}{1}
\expandafter\ifx\csname url\endcsname\relax
  \def\url#1{\texttt{#1}}\fi
\expandafter\ifx\csname urlprefix\endcsname\relax\def\urlprefix{URL }\fi
\providecommand{\bibinfo}[2]{#2}
\providecommand{\eprint}[2][]{\url{#2}}

\bibitem{Ikeda1979}
\bibinfo{author}{Ikeda, K.}
\newblock \bibinfo{title}{Multiple-valued stationary state and its instability
  of the transmitted light by a ring cavity system}.
\newblock \emph{\bibinfo{journal}{Opt. Commun.}} \textbf{\bibinfo{volume}{30}},
  \bibinfo{pages}{257--261} (\bibinfo{year}{1979}).

\bibitem{Hansson2015}
\bibinfo{author}{Hansson, T.} \& \bibinfo{author}{Wabnitz, S.}
\newblock \bibinfo{title}{Frequency comb generation beyond the
  {Lugiato--Lefever} equation: multi-stability and super cavity solitons}.
\newblock \emph{\bibinfo{journal}{J. Opt. Soc. Am. B, JOSAB}}
  \textbf{\bibinfo{volume}{32}}, \bibinfo{pages}{1259--1266}
  (\bibinfo{year}{2015}).

\bibitem{Hansson2016}
\bibinfo{author}{Hansson, T.} \& \bibinfo{author}{Wabnitz, S.}
\newblock \bibinfo{title}{Dynamics of microresonator frequency comb generation:
  models and stability}.
\newblock \emph{\bibinfo{journal}{Nanophotonics}} \textbf{\bibinfo{volume}{5}},
  \bibinfo{pages}{231--243} (\bibinfo{year}{2016}).

\bibitem{Hashemi2024}
\bibinfo{author}{Hashemi, S.~D.} \& \bibinfo{author}{Mittal, S.}
\newblock \bibinfo{title}{Floquet topological dissipative kerr solitons and
  incommensurate frequency combs}.
\newblock \emph{\bibinfo{journal}{Nat. Commun.}} \textbf{\bibinfo{volume}{15}},
  \bibinfo{pages}{1--9} (\bibinfo{year}{2024}).

\bibitem{NLOptics_Boyd}
\bibinfo{author}{Boyd, R.~W.}
\newblock \emph{\bibinfo{title}{Nonlinear Optics, Third Edition}}
  (\bibinfo{publisher}{Academic Press, Inc.}, \bibinfo{address}{USA},
  \bibinfo{year}{2008}), \bibinfo{edition}{3rd} edn.

\bibitem{Pfeiffer2017}
\bibinfo{author}{Pfeiffer, M. H.~P.} \emph{et~al.}
\newblock \bibinfo{title}{Octave-spanning dissipative kerr soliton frequency
  combs in $\text{Si}_{3}\text{N}_{4}$ microresonators}.
\newblock \emph{\bibinfo{journal}{Optica}} \textbf{\bibinfo{volume}{4}},
  \bibinfo{pages}{684} (\bibinfo{year}{2017}).

\end{thebibliography}
%\begin{thebibliography}{}
 %  \input{Main_NH_Comb_arXiv.bbl}
%\end{thebibliography}

\end{document}

	%\preprint{APS/123-QED}
	
	\title{Supplementary Materials\\ Reconfigurable Non-Hermitian Soliton Combs using Dissipative Couplings and Topological Windings}
	
	\author{Seyed Danial Hashemi}
	
	\author{Sunil Mittal}
	\email[Email: ]{s.mittal@northeastern.edu}
	
	\affiliation{Department of Electrical and Computer Engineering, Northeastern University, Boston, MA 02115, USA}
	\affiliation{Institute for NanoSystems Innovation, Northeastern University, Boston, MA, 02115, USA}
	%\date{\today}

\maketitle

%%%%%%%%%%%%%%%%%%%%%%%%%%%%%%%%%%%%%%%%%%%%%%%%%%%%%%%%%%%%%%%%%%%%%%%%%%%%%
% Section
%%%%%%%%%%%%%%%%%%%%%%%%%%%%%%%%%%%%%%%%%%%%%%%%%%%%%%%%%%%%%%%%%%%%%%%%%%%%
\large{\section{Ikeda Map Formalism}}
The formation of dissipative Kerr solitons and coherent optical frequency combs in single and weakly coupled resonators can be efficiently simulated using the Lugiato-Lefever equations (LLEs). However, this formalism relies on the single-mode approximation and the effective Hamiltonian approach, which leads to spurious gain for some of the supermodes of the non-Hermitian array, even when the linear system is completely passive with no gain (see Supplementary Material). Indeed, the transmission and the winding spectrum of the array calculated using a transfer-matrix-based approach show that all the supermodes experience loss because of the dissipative couplings. Nevertheless, depending on the coupling strengths, the hopping phases, and the dissipation strength in link rings, different supermodes experience very different effective losses. Therefore, to simulate the generation of soliton combs in our non-Hermitian array, we use the Ikeda map formalism, which is inherently based on the transfer-matrix approach \cite{Ikeda1979, Hansson2015, Hansson2016, Hashemi2024}.

In the Ikeda map formalism, the electric field $E_{r}^{m}\left(z,\tau\right)$ for a ring number $r$, where $r = \left(1,N \right)$, at a location $z$ along the ring waveguide and at time $\tau = \left(0,\tau_{R}\right)$ is given as 
\begin{eqnarray}
\frac{dE_{r}^{m}\left(z,\mu\right)}{dz} =  i\frac{2\pi}{L_{R}\Omega_{R}}\left(\omega_{p} + \Omega_{R} \mu - \omega_{0\mu} \right) E_{r}^{m}\left(z,\mu\right) - \alpha  E_{r}^{m}\left(z,\mu\right)  \nonumber \\
+ ~i \gamma \frac{1}{\tau_{R}} \int_{0}^{\tau_{R}} d\tau \left|E_{r}^{m}\left(z,\tau\right)\right|^{2} E_{r}^{m}\left(z,\tau\right) e^{i\omega_{\mu} \tau}.
\end{eqnarray}

Here $m$ is the round-trip number such that any time $t = \tau + \left(m-1\right) \tau_{R}$. The Fourier transform of the field $E_{r}^{m}\left(z,\tau\right)$ with respect to the fast time $\tau$ yields $E_{r}^{m}\left(z,\mu\right)$ which is the field in the fast frequency domain indicated by the longitudinal mode index $\mu$. $\gamma = 2 \epsilon_{0} n_{0} n_{2} \omega_{0}$ is the strength of Kerr nonlinearity \cite{NLOptics_Boyd} and $\alpha = \kappa_{in}/v_{g}$ is the propagation loss in the ring waveguides. For a lattice with both nearest and next-nearest couplings, we index the rings such that $r = 1,4,7,..$ refer to the site rings, $r = 2,5,8,..$ refer to the link rings connecting neighboring site rings, and $r = 3,6,9,..$ refer to the link rings connecting next-nearest neighbor site rings. Therefore, for a lattice with $N = 20$ site rings, the array consists of a total of 60 rings when both the nearest and next-nearest couplings are present. We also account for the slightly different length of the link rings compared to those of the site rings.

We assume the individual ring resonators to have an anomalous dispersion such that they generate a bright comb. This dispersion is included in the ring resonance frequencies as 
\begin{equation}
 \omega_{0,\mu} = \omega_{0} + \Omega_{R} ~\mu + \frac{D_{2}}{2} \mu^{2},
\end{equation}
where $\omega_{0}$ is the resonance frequency of the pumped mode $\left(\mu = 0 \right)$, $\Omega_{R} = 2\pi \frac{v_{g}}{L_{R}}$ is the free spectral range (in angular frequency units), $L_{R}$ is the length of the ring resonator, and  $D_{2}$ is the second-order anomalous dispersion which we assume to be $5 \times 10^{-6} \Omega_{R}$. 

To further simplify the simulation framework, we use dimensionless parameters and normalize coordinates such that $z ~\rightarrow ~\frac{z}{L_{R}} = \left\{ 0,1 \right\}$, fast-time $\tau ~\rightarrow ~\frac{\tau}{\tau_{R}} = \left\{0,1 \right\}$, FSR  $\Omega_{R} = 2\pi$, and the group velocity $v_{g} = 1$. Furthermore, the dimensionless fields $E_{r}^{m}\left(z,\tau\right) ~\rightarrow ~\sqrt{\gamma ~L} E_{r}^{m}\left(z,\tau\right)$ such that $\gamma = 1$ effectively. The nonlinear propagation is then written using the dimensionless parameters as
\begin{eqnarray}
    \frac{dE_{r}^{m}\left(z,\mu\right)}{dz} =  i\left(\omega_{p} - \omega_{0} - \frac{D_{2}}{2} \mu^{2} \right) E_{r}^{m}\left(z,\mu\right) - \alpha  E_{r}^{m}\left(z,\mu\right) \nonumber \\ 
    + ~i \int_{0}^{1} d\tau \left|E_{r}^{m}\left(z,\tau\right)\right|^{2} E_{r}^{m}\left(z,\tau\right) e^{i\omega_{\mu} \tau}.
\end{eqnarray}

Both the nearest and the nearest neighbor couplings between the site rings are mediated by link rings. Each such coupling region between the site and the link rings is described as 
\begin{equation}
\begin{pmatrix} E_{r}^{m}\left(z_{c}, \tau \right)\\ E_{r'}^{m}\left(z_{c},\tau \right) \end{pmatrix} = \begin{pmatrix} t &i\kappa \\ i\kappa &t\end{pmatrix} \begin{pmatrix} E_{r}^{m}\left(z_{c},\tau \right)\\ E_{r'}^{m}\left(z_{c},\tau \right) \end{pmatrix}.
\end{equation}
Here $t = t_{N}, t_{NN}$, and $\kappa = \kappa_{N}, \kappa_{NN}$ are the transmission and coupling coefficients for nearest and next-nearest neighbor couplings between rings indexed $r$ and $r'$, respectively. 

The dissipative coupling region of the link rings is modeled as
\begin{equation}
E_{r}^{m}\left(z_{NH}, \tau \right) = t^{NH} ~E_{r}^{m}\left(z_{NH}, \tau \right),
\end{equation}
where $t^{NH} = t^{NH}_{N}, t^{NH}_{NN}$ dictates the transmission/loss in the link ring section for nearest and next-neighbor couplings, respectively. $z_{NH}$ refers to the location of the link rings where we introduce the non-Hermitian coupling waveguide. 

The coupling of the input-output waveguide to the lattice at location $r = 1$ is similarly described, using coupling coefficient $\kappa_{IO}$, such that 
\begin{equation}
 E_{out}^{m}\left(\tau\right) = i\kappa_{IO} ~E_{1}^{m}\left(z=1, \tau \right) + t_{IO} E_{in} \left(\tau \right).
\end{equation}
Here, $E_{in} \left(\tau \right)$ is the input pump field, and $E_{out}^{m}\left(\tau\right)$ is the output field. This field $E_{out}^{m}\left(\tau\right)$ yields the temporal output of the generated frequency comb at any time $t = \left(m-1\right) \tau_{R} + \tau$, and its Fourier transform yields the frequency spectrum $E_{out}\left(\omega\right)$. The coupling coefficient $\kappa_{IO}$ is related to the input-output coupling strength $\kappa_{ex} = \frac{\kappa_{IO}^{2}}{2} \frac{v_{g}}{L_{R}}$, where $v_{g}$ is the group velocity. For our simulations, we chose $\kappa_{ex} = 0.1 J$. Our choice of normalized device parameters, for example, an FSR $\Omega_{R}/\left(2\pi\right) = 250 ~\text{GHz}$, $J/\left(2\pi\right) = 2.5 ~\text{GHz}$, $\kappa_{in}/\left(2\pi\right) = 25 ~\text{MHz}$, $D_{2}/\left(2\pi\right) = 1.250 ~\text{MHz}$ can be achieved using low-loss silicon-nitride resonators \cite{Pfeiffer2017}. For these parameters, the required pump power, $P_{in} = \frac{c}{n_{2} \omega_{0} L_{R}} A_{eff} \left|E_{in} \right|^{2} \simeq 0.8 W$ to generate soliton combs in the nearest-neighbor coupled device, and $P_{in} = \sim 20 ~W$ for the next-nearest-neighbor coupled devices. Nevertheless, the required pump power could be significantly reduced by reducing the inter-resonator coupling strength $J$, that is, increasing the loaded quality factor of the rings and also by optimizing the input-output coupling strength $\kappa_{ex}$.\\
\\

%%%%%%%%%%%%%%%%%%%%%%%%%%%%%%%%%%%%%%%%%%%%%%%%%%%%%%%%%%%%%%%%%%%%%%%%%%%%%
% Section
%%%%%%%%%%%%%%%%%%%%%%%%%%%%%%%%%%%%%%%%%%%%%%%%%%%%%%%%%%%%%%%%%%%%%%%%%%%%

\large{\section{Hamiltonian Formalism}}
Close to a longitudinal mode resonance frequency $\omega_{0,\mu}$ of the site rings, indexed by an integer $\mu$, the linear non-Hermitian coupled-resonator array can be intuitively described using an effective Hamiltonian as (37-39)

\begin{align}
\label{Hamiltonian}
H &= \sum_{r} \omega_{0,\mu} a^{\dag}_{r,\mu} ~a_{r,\mu} - J_{N} e^{+i\phi_{N}} a^{\dag}_{r+1,\mu} ~a_{r,\mu} - J_{N} e^{-\delta_{N}} e^{-i\phi_{N}} a^{\dag}_{r-1,\mu} ~a_{r,\mu} \nonumber \\ 
&- J_{NN} e^{+i\phi_{NN}} a^{\dag}_{r+2,\mu} ~a_{r,\mu} - J_{NN} e^{-\delta_{NN}} e^{-i\phi_{NN}} a^{\dag}_{r-2,\mu} ~a_{r,\mu}. 
\end{align}

Here $a^{\dag}_{r,\mu}$ and $a_{r,\mu}$ are photon creation and annihilation operators at a site $r$ and longitudinal mode (FSR) $\mu$. As discussed in the main text, $J_{N}$ and $J_{N} e^{-\delta_{N}}$ are the coupling strengths for photons moving towards the right (CCW) and left (CW) nearest neighbors, respectively. Similarly, $J_{NN}$ and $J_{NN} e^{-\delta_{NN}}$ are the coupling strengths for the next-nearest neighbors. Note that in this linear Hamiltonian, the super-modes at different longitudinal modes $\mu$ do not interact.

\begin{figure*} [h]
 \centering
 \includegraphics[width=0.98\textwidth]{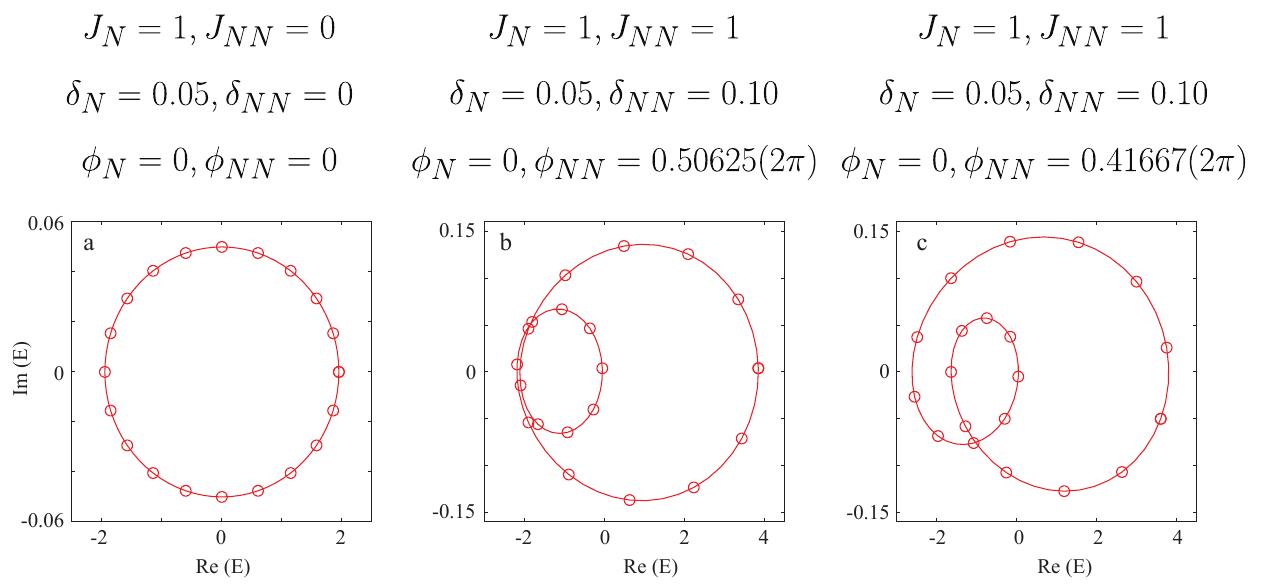}
 \caption{\textbf{Hamiltonian formalism}. \textbf{a-c.} Eigenvalues of the array and topological windings calculated using the Hamiltonian formalism (single-mode approximation). Although the windings are qualitatively similar to those calculated using the transfer-matrix approach (Fig.1 d,j,p), the eigenvalues exhibit a positive imaginary part. The parameters used for these calculations are the same as those reported in the main text. }
 \label{fig:Winding_CMT}
\end{figure*}

The energy eigenvalues for this Hamiltonian are shown in Fig.\ref{fig:Winding_CMT} in the form of topological windings. We find that the windings are qualitatively similar to those observed in Fig.1 of the main text using the transfer-matrix approach. Nevertheless, unlike the windings in Fig.1, the imaginary parts of the eigenvalues of this non-Hermitian Hamiltonian exhibit both positive and negative signs, indicating that some of the supermodes experience gain whereas others experience loss. This is simply a mathematical feature emerging from the effective Hamiltonian formalism because, in the absence of optical nonlinearity, the linear system is completely passive, and none of the supermodes have any gain. To avoid these spurious gain terms, we used the Ikeda map formalism, which is based on the transfer matrix approach, to simulate the generation of optical frequency combs in this non-Hermitian system.

%%%%%%%%%%%%%%%%%%%%%%%%%%%%%%%%%%%%%%%%%%%%%%%%%%%%%%%%%%%%%%%%%%%%%%%%%%%%%
% Section
%%%%%%%%%%%%%%%%%%%%%%%%%%%%%%%%%%%%%%%%%%%%%%%%%%%%%%%%%%%%%%%%%%%%%%%%%%%%

\begin{figure*} [!ht]
 \centering
 \includegraphics[width=0.98\textwidth]{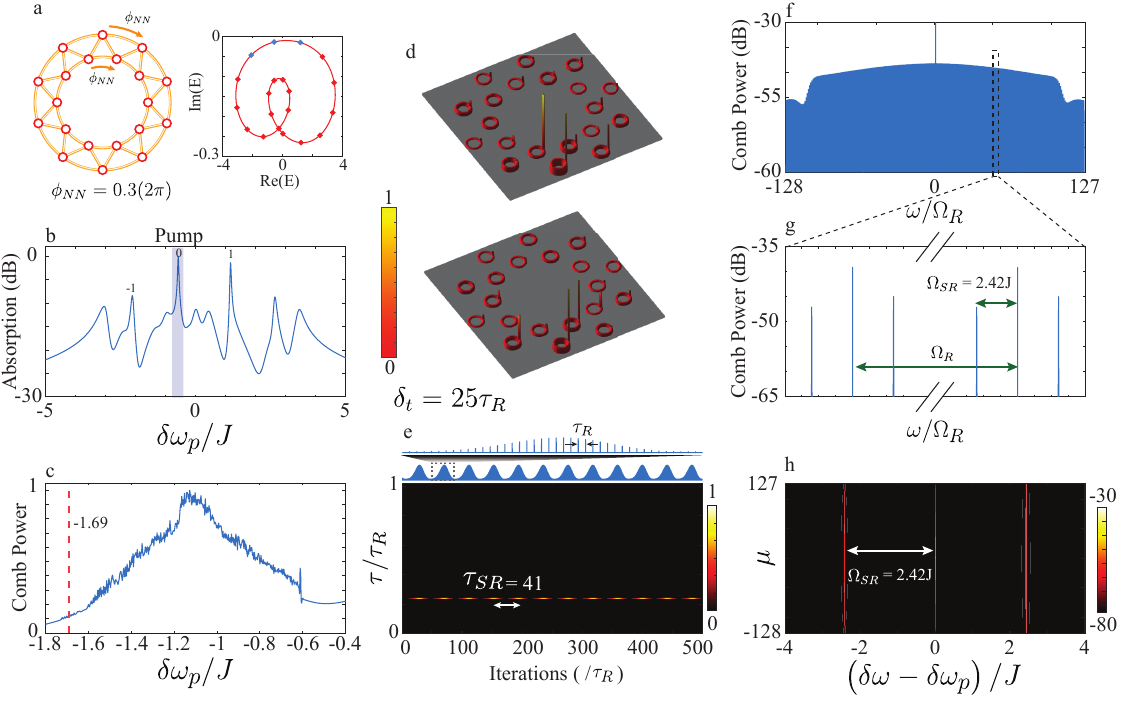}
 \caption{\textbf{Soliton combs in an array with phase $\phi_{NN} = 0.3 (2\pi)$}. \textbf{a.} Topological winding and \textbf{b.} linear absorption spectrum for this choice. \textbf{c.} Generated comb power as a function of pump frequency. The dashed line indicates the pump frequency at which we observe the soliton comb. \textbf{d.} Spatial intensity distribution in the array shows the formation of soliton pulses in rings. \textbf{e.} Temporal output showing bursts of pulses. \textbf{f-h.} Generated comb spectrum showing the oscillation of only three supermodes in each FSR $\mu$. }
 \label{fig:NNN_Phase_3}
\end{figure*}

\large{\section{Soliton Combs for Other Parameters}}
In the main text, we presented results for only a few parameter choices for the ring resonator array. The array can support soliton combs in other parameter regimes as well. As an example, in Fig.\ref{fig:NNN_Phase_3}, we show the results for a soliton comb generated in an array with $J_{N} = J_{NN}$, $\phi_{N} = 0$, $\phi_{NN} = 0.3 (2\pi)$, $t_{NH}^{N} = 0.95$ and $t_{NH}^{NN} = 0.9$. Except for the phase $\phi_{NN}$, these parameters are the same as those of Figs. 3 and 4 of the main text. In this case, the 2D complex energy plane exhibits a double winding, similar to those of Figs. 2 and 3. However, we observe a nested comb with only three oscillating supermodes in each FSR $\mu$. 

\begin{figure*} [h]
 \centering
 \includegraphics[width=0.98\textwidth]{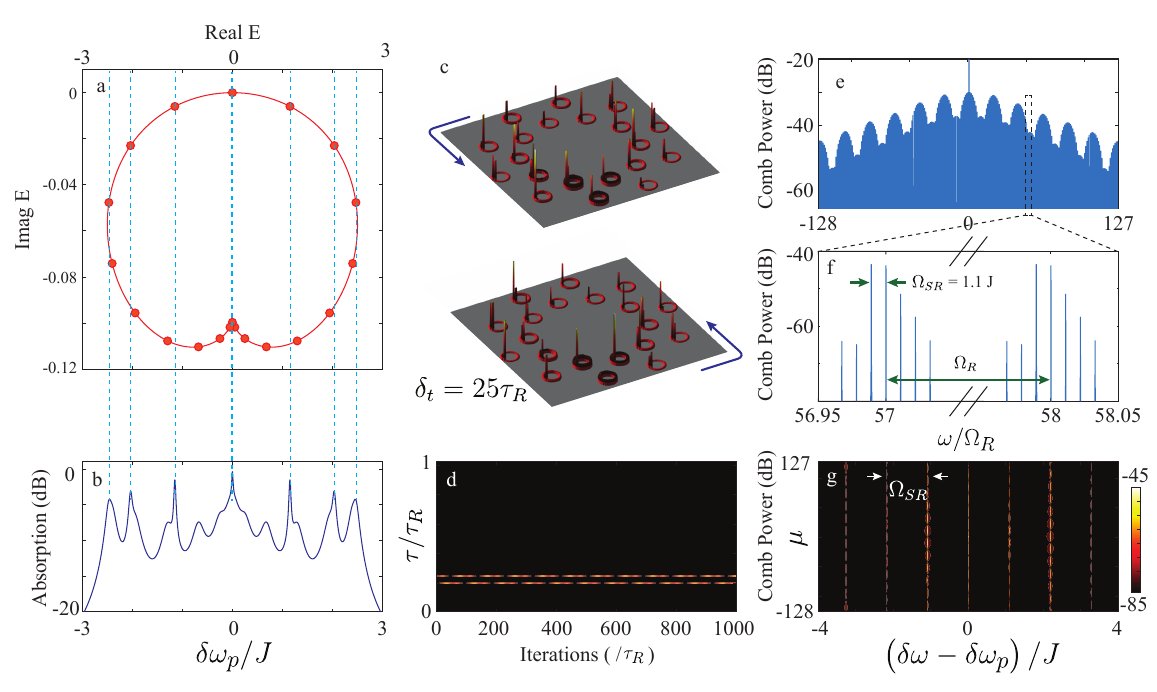}
 \caption{\textbf{Nested soliton combs in an array with $J_{NN} = 0.5 J_{N}$, $\phi_{N} = 0$, $\phi_{NN} = 0.25 (2\pi)$, $t_{NH}^{N} = 0.95$ and $t_{NH}^{NN} = 0.95$.} \textbf{a.} Topological winding in the complex plane and, \textbf{b.} linear absorption spectrum. \textbf{c} Spatio-temporal intensity distribution in the array showing the presence of two solitons in each ring. \textbf{d.} Temporal output where the presence of two solitons in the rings is also observed in the fast time domain $\tau$. \textbf{e-g.} Generated comb spectrum showing dominant oscillation of seven supermodes in each FSR.}
 \label{fig:NNN_7_Mode}
\end{figure*}

We expect to observe similar reconfigurability for different parameter choices as well. For example, with parameters $J_{NN} = 0.5 J_{N}$, $\phi_{N} = 0$, $\phi_{NN} = 0.25 (2\pi)$, $t_{NH}^{N} = 0.95$ and $t_{NH}^{NN} = 0.95$, we observe a nested comb with predominantly seven oscillating supermodes in each FSR (Fig.\ref{fig:NNN_7_Mode}). In this comb, there are two solitons in each ring, and both are phase-locked across the rings of the array. Nevertheless, we emphasize that not every choice of array parameters is expected to generate nested soliton combs. In particular, the generation of soliton combs in the non-Hermitian array will depend on the dispersion and dissipation of the supermodes, which will need to be engineered to achieve a given number of comb lines and to prevent nonlinear mixing between desired and undesired modes.  \\

%%%%%%%%%%%%%%%%%%%%%%%%%%%%%%%%%%%%%%%%%%%%%%%%%%%%%%%%%%%%%%%%%%%%%%%%%%%%%
% Section
%%%%%%%%%%%%%%%%%%%%%%%%%%%%%%%%%%%%%%%%%%%%%%%%%%%%%%%%%%%%%%%%%%%%%%%%%%%%

\begin{figure*} [ht]
 \centering
 \includegraphics[width=0.98\textwidth]{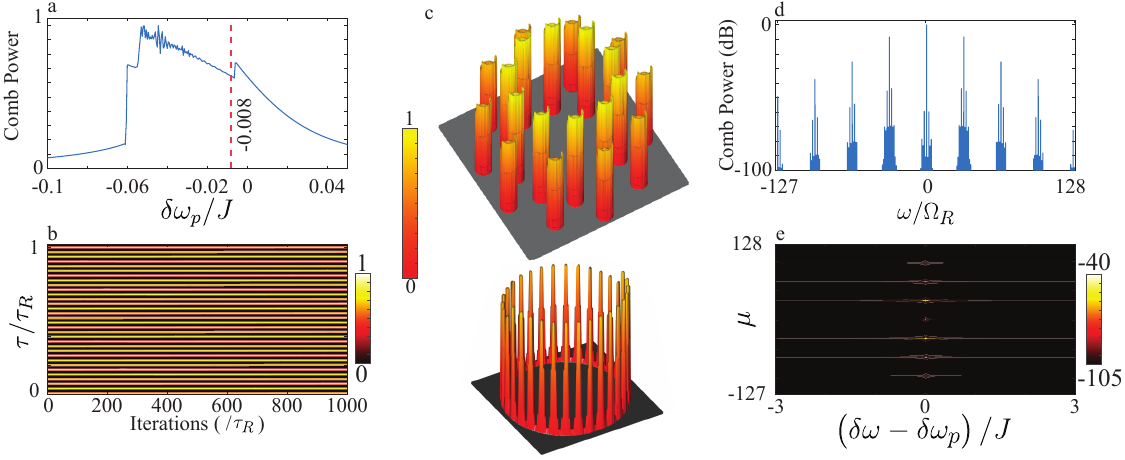}
 \caption{\textbf{Turing rolls in the super-ring resonator with only nearest neighbor couplings.} \textbf{a.} Pump power in the input-output ring. \textbf{b.} Temporal output shows a series of pulses repeating at a time scale much smaller than the round-trip time $\tau_{R}$ of the single rings. \textbf{c.} Intensity distribution in the lattice shows nearly uniform occupation of all the rings. The inset shows Turing roll pattern in a single ring. The Turing rolls are also phase-locked across all rings. \textbf{d.} Generated comb spectrum showing oscillation of only a small subset of FSRs. \textbf{e.} Comb spectrum as a function of fast frequency $\mu$ and slow frequency $\left(\delta\omega-\delta\omega_{p}\right)$ shows oscillation of only a single supermode.}
 \label{fig:Turing}
\end{figure*}

\large{\section{Turing Rolls in Super-Ring Resonator}}
In addition to the soliton combs, where there is a single pulse propagating, ring resonators also support coherent Turing roll patterns where there are multiple pulses propagating in the ring. We have also observed such Turing roll patterns in our non-Hermitian coupled resonator array. In Fig.\ref{fig:Turing}, we show results for observed Turing rolls in an array with only nearest neighbor couplings. We pump the same supermode as we did to achieve soliton combs in Fig. 2 of the main text, but we lower the pump field amplitude $E_{in} = 0.012$. At $\delta\omega_{p} = -0.008 ~J$, we find that each site ring of the array hosts Turing rolls. Furthermore, similar to the solitons, the Turing rolls are phase-locked across all rings. 

In the time domain, the output consists of a series of pulses but with a repetition time much smaller than $\tau_{R}$. In the frequency domain, we observe the oscillation of only a few single-ring longitudinal modes, separated by $ 32 \Omega_{R}$. More importantly, within each FSR, we observe the oscillation of only a single supermode. This coherent comb state can be compared to that of Fig.4 of the main text, where we also observed the oscillation of a single supermode but across all FSRs.\\

%%%%%%%%%%%%%%%%%%%%%%%%%%%%%%%%%%%%%%%%%%%%%%%%%%%%%%%%%%%%%%%%%%%%%%%%%%%%%
% Section
%%%%%%%%%%%%%%%%%%%%%%%%%%%%%%%%%%%%%%%%%%%%%%%%%%%%%%%%%%%%%%%%%%%%%%%%%%%%

\begin{figure*}[ht]
 \centering
 \includegraphics[width=0.98\textwidth]{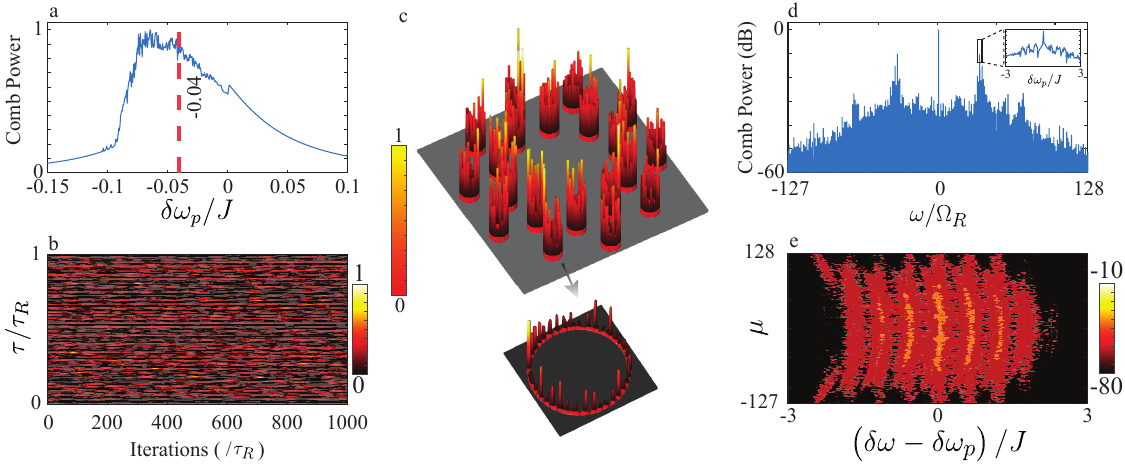}
 \caption{\textbf{Chaotic combs in super-ring resonator with only nearest neighbor couplings.} \textbf{a.} Pump power in the input-output ring. \textbf{b.} Temporal output shows no periodic patterns. \textbf{c.} Intensity distribution in the lattice occupies all the rings, but the intensity distribution across the rings is not uniform. The inset shows intensity distribution within a single ring. Different rings have different intensity distributions. \textbf{d,e.} The generated comb spectrum is not smooth, indicating a lack of coherence. Nevertheless, the comb spectrum shows the oscillation of multiple supermodes. Their curvature in \textbf{e.} highlights their linear dispersion, which has not been compensated by the nonlinear dispersion.}
 \label{fig:Chaotic}
\end{figure*}

\large{\section{Chaotic Combs in Super-Ring Resonator}}
Similar to single-ring Kerr combs, most pump frequencies generate chaotic or modulation instability combs in our ring resonator array. The generated chaotic comb at one such frequency $\delta \omega_{p} = -0.04 ~J$ is shown in Fig.\ref{fig:Chaotic}. This array, with only nearest neighbor couplings, has the same parameters as that of Fig.2 of the main text. In this chaotic regime, we do not observe any self-organization or self-synchronization of pulses across different site rings (Fig.\ref{fig:Chaotic}c). Even more so, we do not observe any soliton-like pulses in individual rings. The comb spectrum is not smooth (Fig.\ref{fig:Chaotic}d). Nevertheless, it does exhibit the oscillation of multiple supermodes (Fig.\ref{fig:Chaotic}e). The curvature of the comb lines indicates no cancellation of linear resonator dispersion by the Kerr nonlinearity.  

%%%%%%%%%%%%%%%%%%%%%%%%%%%%%%%%%%%%%%%%%%%%%%%%%%%%%%%%%%%%%%%%%%%%%%%%%%%%%
% bibliography
%%%%%%%%%%%%%%%%%%%%%%%%%%%%%%%%%%%%%%%%%%%%%%%%%%%%%%%%%%%%%%%%%%%%%%%%%%%%%
\newpage

\bibliography{Suppl_NH_Comb_arXiv.bbl}

%\bibliographystyle{NatureMag}
%\bibliography{FC_Biblio, Topo_Biblio, NH_Biblio}